\documentclass[iop,numberedappendix]{emulateapj}

\usepackage{url}
\usepackage{enumerate}
\usepackage{natbib}

\newcommand{\degree}{$^{\circ}$}
\newcommand{\IRAS}{\textit{IRAS}}

\newcommand{\Ehk}{$E(H - K_{\rm s})$}
\newcommand{\Ebv}{$E(B-V)$}
\newcommand{\Ejh}{$E(J-H)$}
\newcommand{\Ejk}{$E(J-K_{\rm s})$}
\newcommand{\av}{$A_V$}

\newcommand{\nh}{$N_{\rm H}$}

\newcommand{\nhoverejk}{$N_{\rm H}/E(J-K_{\rm s})$}
\newcommand{\Tone}{15.3~$\pm$~0.4~K}
\newcommand{\rectno}{7}
\newcommand{\Tonebav}{15.2~$\pm$~0.2}
\newcommand{\epsilonc}{249.0~$\pm$~5.0} 
\newcommand{\epsiloncamp}{230.3~$\pm$~7.2} 

\slugcomment{Accepted for publication in the Astrophysical Journal}

\begin{document}

\shorttitle{Submillimeter Dust Opacity}

\shortauthors{Martin et al.}


\title{Evidence for Environmental Changes in the Submillimeter Dust Opacity}


\author{Peter~G.~Martin,\altaffilmark{1}
        Arabindo~Roy,\altaffilmark{1}
        Sylvain~Bontemps,\altaffilmark{2}
        Marc-Antoine~Miville-Desch\^enes,\altaffilmark{3,1}
	Peter~A.~R.~Ade,\altaffilmark{4}
        James~J.~Bock,\altaffilmark{5,6}
	Edward~L.~Chapin,\altaffilmark{7}
	Mark~J.~Devlin,\altaffilmark{8}
	Simon~R.~Dicker,\altaffilmark{8}
	Matthew~Griffin,\altaffilmark{4}
	Joshua~O.~Gundersen,\altaffilmark{9}
        Mark~Halpern,\altaffilmark{7}
        Peter~C.~Hargrave,\altaffilmark{4}
	David~H.~Hughes,\altaffilmark{10}
	Jeff~Klein,\altaffilmark{8}
	Gaelen~Marsden,\altaffilmark{7}
        Philip~Mauskopf,\altaffilmark{4}
	Calvin~B.~Netterfield,\altaffilmark{11,12}
        Luca~Olmi,\altaffilmark{13,14}
	Guillaume~Patanchon,\altaffilmark{15}
	Marie~Rex,\altaffilmark{8}
        Douglas~Scott,\altaffilmark{7}
	Christopher~Semisch,\altaffilmark{8}
	Matthew~D.~P.~Truch,\altaffilmark{8}
	Carole~Tucker,\altaffilmark{4}
        Gregory~S.~Tucker,\altaffilmark{16}
	Marco~P.~Viero,\altaffilmark{17}
	Donald~V.~Wiebe\altaffilmark{7}}

\altaffiltext{1}{Canadian Institute for Theoretical Astrophysics, University of Toronto, 60 St. George Street, Toronto, ON M5S~3H8, Canada}

\altaffiltext{2}{Observatoire de Bordeaux, BP 89, F-33270 Floirac, France} 

\altaffiltext{3}{Institut d'Astrophysique Spatiale, CNRS (UMR8617) Universit\'e Paris-Sud 11, B\^atiment 121, Orsay, France}

\altaffiltext{4}{Department of Physics \& Astronomy, Cardiff University, 5 The Parade, Cardiff, CF24~3AA, UK}

\altaffiltext{5}{Jet Propulsion Laboratory, Pasadena, CA 91109-8099}

\altaffiltext{6}{Observational Cosmology, MC 59-33, California Institute of Technology, Pasadena, CA 91125}

\altaffiltext{7}{Department of Physics \& Astronomy, University of British Columbia, 6224 Agricultural Road, Vancouver, BC V6T~1Z1,Canada}

\altaffiltext{8}{Department of Physics and Astronomy, University of Pennsylvania, 209 South 33rd Street, Philadelphia, PA 19104}

\altaffiltext{9}{Department of Physics, University of Miami, 1320 Campo Sano Drive, Carol Gables, FL 33146}

\altaffiltext{10}{Instituto Nacional de Astrof{\'i}sica {\'O}ptica y Electr{\'o}nica (INAOE), Aptdo. Postal 51 y 72000 Puebla, Mexico}

\altaffiltext{11}{Department of Astronomy \& Astrophysics, University of Toronto, 50 St. George Street, Toronto, ON M5S~3H4, Canada}

\altaffiltext{12}{Department of Physics, University of Toronto, 60 St. George Street, Toronto, ON M5S~1A7, Canada}

\altaffiltext{13}{INAF, Osservatorio Astrofisico di Arcetri, Largo E. Fermi 5, I-50125, Italy}

\altaffiltext{14}{University of Puerto Rico, Rio Piedras Campus, Physics Dept., Box 23343, UPR station, San Juan, Puerto Rico}

\altaffiltext{15}{Laboratoire APC, 10, rue Alice Domon et L{\'e}onie Duquet 75205 Paris, France}

\altaffiltext{16}{Department of Physics, Brown University, 182 Hope Street, Providence, RI 02912}

\altaffiltext{17}{Department of Astronomy, MC 367-17, California Institute of Technology, Pasadena, CA 91125, U.S.A.}



\begin{abstract} 
 
The submillimeter opacity of dust in the diffuse interstellar medium in
the Galactic plane has been quantified using a pixel-by-pixel
correlation of images of continuum emission with a proxy for column
density.  We used multi-wavelength continuum data: three BLAST bands at
250, 350, and 500~\micron\ and one \IRAS\ at 100~\micron.  The proxy is
the near-infrared color excess, \Ejk, obtained from 2MASS. Based on
observations of stars, we show how well this color excess is correlated
with the total hydrogen column density for regions of moderate
extinction.  The ratio of emission to column density, the emissivity, is
then known from the correlations, as a function of frequency.  The
spectral distribution of this emissivity can be fit by a modified
blackbody, whence the characteristic dust temperature $T$ and the
desired opacity $\sigma_{\rm e}(1200)$ at 1200 GHz or 250~\micron\ can
be obtained.  We have analyzed 14 regions near the Galactic plane toward
the Vela molecular cloud, mostly selected to avoid regions of high
column density ($N_{\rm {H}} > 10^{22}$~cm$^{-2}$) and small enough to
ensure a uniform dust temperature.  We find $\sigma_{\rm e}(1200)$ is
typically 2 to $4 \times 10^{-25}$~cm$^2$~H$^{-1}$ and thus about 2 to 4
times larger than the average value in the local high Galactic latitude
diffuse atomic interstellar medium.  This is strong evidence for grain
evolution. There is a range in total power per H nucleon absorbed (and
re-radiated) by the dust, reflecting changes in the strength of the
interstellar radiation field and/or the dust absorption opacity.  These
changes in emission opacity and power affect the equilibrium $T$, which
is typically 15 K, colder than at high latitudes. Our analysis extends,
to higher opacity and lower temperature, the trend of increasing
$\sigma_{\rm e}(1200)$ with decreasing $T$ that was found at high
latitudes.  The recognition of changes in the emission opacity raises a
cautionary flag because all column densities deduced from dust emission
maps, and the masses of compact structures within them, depend inversely
on the value adopted.

\end{abstract}

\keywords{Balloons -- dust, extinction -- evolution -- Infrared: ISM --
  ISM: structure -- Submillimeter: ISM}


\section{Introduction}

Observations of submillimeter dust emission are a prime means for
determining masses in the interstellar medium, including in compact
clumps and cores of star-forming regions.
However, there remains considerable systematic uncertainty because the
opacity $\sigma_{\rm e}$, the dust emission cross-section per H nucleon,
is not well known.
The value of $\sigma_{\rm e}$ is best determined for the diffuse atomic
interstellar medium at high Galactic latitude
\citep{boulanger1996,abergelDd2011}.
The main goal of this paper is to quantify $\sigma_{\rm e}$ empirically
in different environments near the Galactic plane, characterized by
higher column and spatial density and at least in part molecular.

Detailed models of interstellar grains
\citep{dwek1997,lidraine2001,compiegne2011} are best constrained in the
local diffuse and largely atomic interstellar medium. 
The value of $\sigma_{\rm e}$ can be calculated for a given model of
dust grains, depending upon the composition and grain structure but
fortunately not so strongly on size or shape.  This calculation involves
the emission cross section per gram of dust, $\kappa_\nu$, and the
dust-to-gas mass ratio, $r$.  Thus these models are further constrained
by and/or checked for the consistency of the product $r \kappa_\nu$ with
the empirical value of $\sigma_{\rm e}$ (Equation~(\ref{cross})) at high
Galactic latitude.

The value of the submillimeter opacity is likely to change with
environment, through differences in composition, structure, and even
dust temperature.  Certainly there are changes in the
optical-ultraviolet extinction curve that indicate and even quantify
some aspects of dust evolution (\citealp{kimmartin1996} and references
therein).  Pioneering work on NGC~7023, correlating submillimeter
emission with a measure of ultraviolet extinction, demonstrated
empirically that the opacity might be higher by a factor of two in that
denser environment \citep{hildebrand1983}. However, the uncertainty was
much too large (a factor three or four) to be definitive.

Where, when, and how evolution happens are unknown, but it seems
plausible that in denser and colder regions, where dust grains acquire
ice mantles, the dust is more susceptible to aggregation on a relevant
time scale.
Theoretical dust models can be used to explore the complexities of the
observable consequences (extinction, emission) of dust evolution in
denser environments (e.g., \citealp{ossenkopf1994,ormel2011}; see also
Section~\ref{insight}) but the evolution is undoubtedly too complex for
precise environment-specific predictions {\it ab initio}.  Thus we take
an empirical approach to quantifying the opacity in different
environments of higher column and spatial density.

The basis of our study is a careful correlation of the emission in
submillimeter dust continuum images obtained by the Balloon-borne Large
Aperture Submillimeter Telescope (BLAST), together with images from
\IRAS, with a proxy for column density, the near-infrared color excess
that is obtained from analysis of 2MASS\footnote{The Two Micron All Sky
  Survey (2MASS) is a joint project of the University of Massachusetts
  and the Infrared Processing and Analysis Center/California Institute
  of Technology, funded by the National Aeronautics and Space
  Administration and the National Science Foundation.} data.  The slopes
of these correlations are emissivities.
BLAST mapped simultaneously at 250, 350, and
500~\micron\ \citep{pascale2008} and so together with
\IRAS\ 100~\micron\ data the spectral coverage is quite good in the
region of the peak of the spectral energy distribution (SED) of the
emissivities, allowing the characteristic dust temperature $T$ to be
constrained via a modified blackbody fit.  Knowing $T$ is essential for
recovering the dust opacity from the dust emissivity.  Accompanying
goals are to assess the uncertainties, to acknowledge explicitly some
``known unknowns,'' to show how our results extend the range of
environments in which there is a reasonable calibration of the opacity,
and to explore trends between variations in $T$ and $\sigma_{\rm e}$.

Our paper is organized deliberately to isolate the steps in our approach
and to identify potential sources of systematic error.
In Section~\ref{dustemission} we briefly discuss mechanisms of diffuse
emission and the quantitative relationships of dust emissivity and
opacity to mass estimates from submillimeter continuum emission.
We introduce the BLAST imaging in Section~\ref{imaging} and make our
first estimate of $T$ from the relative SED.
In Section~\ref{extinction} we discuss color excess measured with 2MASS,
a surrogate for column density.
The fundamental correlation of emission with color excess is presented
in Section~\ref{correlation}, where we then use the SED of the
emissivities to obtain the amplitude and a second measure of $T$,
corroborating the first.
In Section~\ref{sec:hydrogen} we determine the ratio of hydrogen column
density to infrared color excess, \Ejk, from stellar data.
This allows us in Section~\ref{results} to determine the desired
opacity $\sigma_{\rm e}$ from the above amplitudes and temperatures
and to compute other important parameters like $P$, the power emitted
(absorbed) per H nucleon.  We assess the errors and in
Appendix~\ref{abeta} the impact of possible systematic changes in the
spectral dependence of the opacity with $T$.  We show that the
parameters $T$, $\sigma_{\rm e}$, and $P$ change significantly from
region to region.
In Section~\ref{analysis} we discuss the systematic interrelationships
among these changes, bringing in for comparison other estimates of
opacity drawing on a brief summary of the literature in
Appendix~\ref{literature}.  We comment on underlying uncertainties,
range of applicability, and efforts at theoretical modeling.
Finally, Section~\ref{sec:conclusion} gives our conclusions and
anticipates future work.

\section{Submillimeter Dust Emission}\label{dustemission}

BLAST maps of thermal dust emission measure surface brightness $I_\nu$
(MJy sr$^{-1}$) and hence, for optically-thin emission, the mass column
density of the dust, $M_{\rm {d}}$, which is a fraction $r$ (the
dust-to-gas mass ratio) of the total mass column density.  Thus, more
technically,
\begin{equation}
I_\nu = M_{\rm {d}} \kappa_\nu  B_{\nu}(T) =  r \mu m_{\rm{H}}
N_{\rm{H}}\kappa_\nu  B_{\nu}(T) = \tau_\nu  B_{\nu}(T),
\label{emissmass}
\end{equation}
where $\kappa_\nu$ is the dust mass absorption (or emission) coefficient
(cm$^2$g$^{-1}$), often called the opacity, $B_{\nu}(T)$ is the Planck
function for dust temperature $T$, $\mu$ is the mean molecular weight,
$N_{\rm{H}}$ is the total H column density (H in any form), and
$\tau_\nu$ is the optical depth of the column of material. If $T$
changes along the line of sight, then Equation~(\ref{emissmass}) has to be
generalized appropriately, but obviously the analysis is more
straightforward if strong $T$ gradients are avoided, as we will do.

For a (small) region of uniform $T$ the variation in brightness in a map
tracks changes in optical depth along the different lines of sight.  In
fact, as Equation~(\ref{emissmass}) shows, a map of optical depth could be
obtained directly by dividing the image by the Planck function
$B_{\nu}(T)$ if the appropriate dust temperature can be obtained.
However, because the ``zero point'' of the BLAST emission maps is not
known we do not use this direct approach here.

The emissivity of a column of interstellar material,
\begin{equation}
\epsilon_{\rm e}(\nu) \equiv I_\nu/N_{\rm{H}},
\label{emissH}
\end{equation}
is an observable prerequisite to quantifying the desired opacity.  By
analogy with Equation~(\ref{emissH}), for a column of interstellar material
characterized by its extinction, we can define
\begin{equation}
\epsilon_{\rm c}(\nu) \equiv I_\nu/E(J-K_{\rm s}),
\label{emisscolour}
\end{equation}
where \Ejk\ is the color excess from differential extinction between the
$J$ and $K_{\rm s}$ passbands (of 2MASS); $\epsilon_{\rm c}(\nu)$ is the
observable addressed below.

The opacity $\sigma_{\rm e}(\nu)$ of the interstellar material medium
is ultimately a property of the grain material as can be seen from the
relationships
\begin{equation}
\sigma_{\rm e}(\nu) \equiv \tau_\nu/N_{\rm H} = \epsilon_{\rm e}(\nu)/B_{\nu}(T) =
\mu m_{\rm H} r \kappa_\nu.
\label{cross}
\end{equation}
Quantifying the opacity also requires knowledge of $T$, which can be
obtained from the multi-wavelength SED of the emissivities as described
below.

In what follows we parameterize the spectral dependence as $\kappa_\nu =
\kappa_0 (\nu/\nu_0)^\beta$, with fiducial frequency $\nu_0 = 1200$~GHz
($\lambda_0 = 250$~\micron) and $\beta = 1.8$, to compare directly with
the value of $\sigma_{\rm e}(1200)$ for the diffuse interstellar medium
given by \citet{boulanger1996} and \citet{abergelDd2011}.

For identifiable objects in a map (peaks of emission) the integral of
$I_\nu$ over source size gives a flux density.  When flux density is
integrated over $\nu$ across the entire SED, and the distance is known,
a luminosity is obtained.  One does not have to know the opacity to
interpret this observable; however, both opacity and temperature are
required to obtain the mass of the (compact) object.
By analogy, a related quantity for diffuse emission is the power per H
emitted by dust grains (equal to the absorbed power), computed by
integrating the emissivity over $\nu$:
\begin{equation}
P =   \int  4 \pi  \epsilon_{\rm e}(\nu) d\nu = \int  4 \pi \sigma_{\rm e}(\nu) B_\nu(T) d\nu.
\label{power}
\end{equation}
Again this can be computed from the observable set of $\epsilon_{\rm
  e}(\nu)$ values, without knowledge of the opacity.  It can also be
seen from our parameterization of the frequency dependence of the
opacity that $P$ varies as $\sigma_{\rm e}(1200) T^{4+\beta}$.

For the diffuse atomic high latitude interstellar medium
\citep{abergelDd2011}, typical values are $\beta = 1.8$, $T_0 =
17.9$~K, $\sigma_{\rm e}(1200) \equiv \sigma_0 = 1.0 \times
10^{-25}$~cm$^2$~H$^{-1}$ ($r\kappa_0 = 0.043$~cm$^2$~gm$^{-1}$) and
$P_0 = 3.8 \times 10^{-31}$~W~H$^{-1}$.
Perhaps more memorable, expressed in solar units this power is close
to unity: $P_0/m_{\rm H} \sim 1.2~$L$_\odot$/M$_\odot$.
Note that $P$ is not solely a property of the dust; because the
thermally-emitting dust is in radiative equilibrium, $P$ depends
directly on the strength of the interstellar radiation field in which
the dust is immersed.
In the high latitude analysis, the opacity is calibrated through
empirical correlation of dust emission with atomic hydrogen column
density. It is found, surprisingly, that the opacity is not constant
even in that high latitude environment, that one would expect to be
simple.  Our new study extends such studies to regions that have higher
column density and are at least partially molecular.


\section{Submillimeter Observations: BLAST Imaging of Vela}\label{imaging}

\begin{figure}
\includegraphics[width=\linewidth]{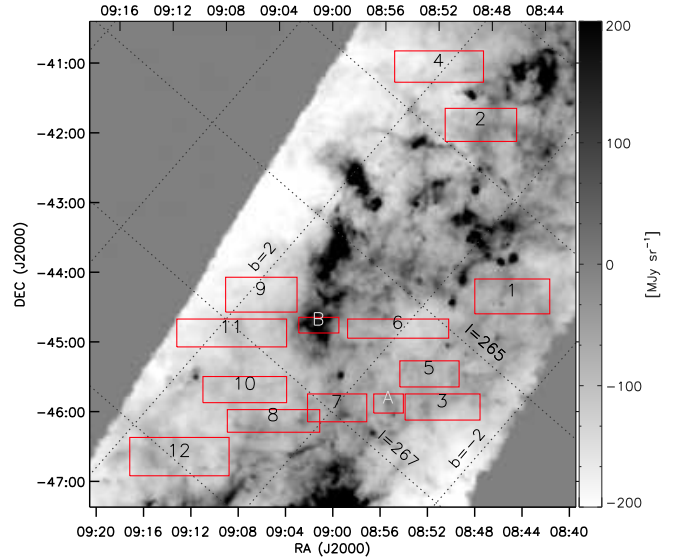} 
\caption{BLAST 250~\micron\ dust emission map spanning the Vela
  Molecular Ridge.  This is projected on the grid of the 2MASS-based
  color-excess map in Figure~\ref{fig:extinction}, at a degraded
  resolution and 1\farcm5 per pixel.  Outlined are several small
  rectangular regions, aligned roughly along the scans, which are
  analyzed independently.  These were selected to avoid gradients in
  dust temperature, compact sources, and (molecular) regions of the
  highest column density.}
\label{wide}
\end{figure}

With BLAST \citep{pascale2008,truch2008,truch2009} we surveyed
50~deg$^2$ in Vela for 10.6~hours during the December 2006 flight
\citep{netterfield2009}.  BLAST06 produced diffraction-limited images
with resolutions (full width at half maximum: FWHM) of 36\arcsec,
42\arcsec, and 60\arcsec\ at 250, 350, and 500~\micron, respectively
\citep{truch2009}.  
For the pixel-pixel correlations with the color excess maps
(Section~\ref{correlation}), we have convolved the BLAST images to
$2\farcm1$ resolution and regridded to $1\farcm5$ pixels.  The
250~\micron\ map used is shown in Figure~\ref{wide}.

\begin{figure}
\includegraphics[width=\linewidth]{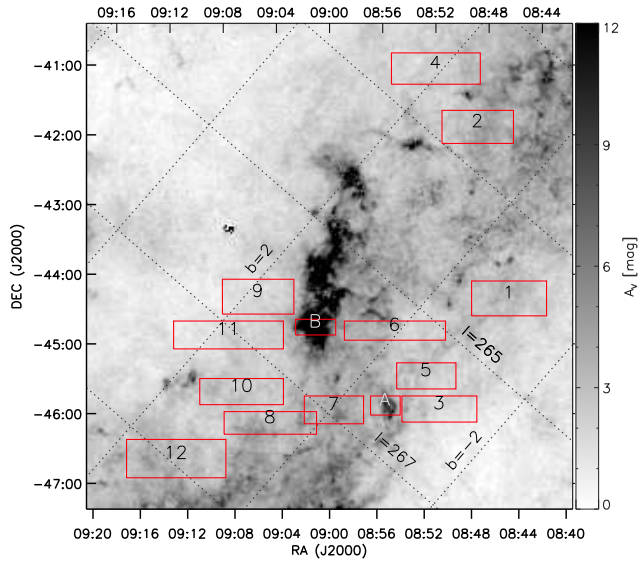} 
\caption{Like Figure~\ref{wide} but for the column density from
  extinction, the color excess \Ejk. The intensity scale has been
  labeled in $A_V$, which is perhaps more intuitive, but as discussed in
  Section~\ref{extinction} this conversion is only approximate and should
  not be the basis for quantitative calculations.}
\label{fig:extinction}
\end{figure}

The observations were performed by scanning the telescope in azimuth at
a speed of $0\fdg2$~s$^{-1}$.  The combination of high scan speed and
low $1/f$ knee, together with the multiple cross-linking and common-mode
removal in the map-maker SANEPIC \citep{pat08} retains diffuse low
spatial frequency emission (the DC level is removed, however).
Ideally the cross-linking scans would have been orthogonal, but solar
position constraints resulted in a small angle range, with scans being
oriented roughly along constant declination.  Consequently small drifts
in the baseline produced a low spatial frequency undulation in the
cross-scan direction, readily apparent by comparing results from
different independent map-makers.  We have examined rectangular
subregions of sufficiently small dimension in declination to mitigate
any potential effects of this artefact (much taller regions were, in
fact, examined too and the results were found to be robust).  

The subregions examined are shown in Figures~\ref{wide} and
\ref{fig:extinction} and specified in Table~\ref{tab1}.  The typical
length of a rectangle, $1\degr$, corresponds to 12~pc at the distance
of the Vela Molecular Ridge, 700~pc \citep{murphy1991}.  The Galactic
latitude range covered by the 
centers of the different rectangles is $-1\degr$ to $+2\degr$.  The
maximum separation in longitude is about $7\degr$ (85~pc).  The
different rectangles sample distinct environments.
 
\begin{deluxetable}{cccccccccc}
\tablewidth{0pt}
\small
\setlength{\tabcolsep}{0.02in}
\tablecaption{Rectangular regions analyzed}
\tablehead{
\colhead{ID}&
\colhead{$\alpha$ (J2000)}&
\colhead{$\delta$ (J2000)}&
\colhead{$l$}&
\colhead{$b$}&
\colhead{Size}\\
\colhead{}&
\colhead{h:m:s}&
\colhead{d:m:s}&
\colhead{$^\circ$}&
\colhead{$^\circ$}&
\colhead{\arcmin\ \, $\times$\, \, \arcmin}\
}
       1&8:45:39.5&-44:24:22.9&264.061&-0.868& 64.5~$\times$ 30.0\\
       2&8:48:38.6&-41:58:41.9&262.511& 1.078& 61.5~$\times$ 28.5\\
       3&8:51:00.2&-46:02:22.4&265.923&-1.166& 64.5~$\times$ 22.5\\
       4&8:51:58.6&-41:08:50.7&262.264& 2.087& 76.5~$\times$ 27.0\\
       5&8:52:04.5&-45:34:11.9&265.680&-0.722& 51.0~$\times$ 22.5\\
       6&8:54:42.2&-44:55:47.2&265.486& 0.043& 87.0~$\times$ 16.5\\
       7&8:59:38.4&-46:03:41.7&266.908&-0.036& 51.0~$\times$ 24.0\\
      8&9:04:59.1&-46:15:16.4&267.666& 0.529& 79.5~$\times$ 19.5\\
      9&9:05:48.6&-44:25:41.8&266.407& 1.860& 61.5~$\times$ 30.0\\
      10&9:07:14.3&-45:47:51.4&267.590& 1.126& 72.0~$\times$ 22.5\\
      11&9:08:16.0&-44:58:07.5&267.101& 1.818& 94.5~$\times$ 24.0\\
      12&9:12:53.4&-46:42:54.4&268.928& 1.216& 85.5~$\times$ 33.0\\
\hline\hline\\
      A&8:55:28.1&-46:00:21.9&266.394&-0.550& 25.5~$\times$ 16.5\\ 
      B&9:01:12.0&-44:53:13.3&266.204& 0.943& 34.5~$\times$ 13.5\\

\label{tab1}
\end{deluxetable}

To avoid areas with strong gradients in dust temperature, we
concentrated on relatively diffuse regions of moderate brightness,  
without compact sources, not the high column (and spatial) density
molecular clouds where conditions might be less uniform along the line
of sight and where the uncertainties in determining the color excess
\Ejk\ are larger.

However, suspending our caution, we examined in addition two prominent
higher column density regions traced even in $^{13}$CO
\citep{yamaguchi1999}; see rectangles A and B.\footnote{Rectangle B is
  in the main Vela~C cloud and unavoidably contains some cold compact
  sources that, although peaks in column density, might potentially produce extra
  scatter in the correlation with the color excess map, as discussed below
  in Section~\ref{correlation}.}
\citet{netterfield2009} derived $\sigma_{\rm e}(1200)$ for the clump
in rectangle A by comparing the integrated BLAST emission with the
mass derived from CO observations \citep{yamaguchi1999}, providing an
independent check on both approaches.

Note that for absolute measures of column density, pixel by pixel, we
would have to restore the zero point (DC level) of the BLAST maps, as we
did for the Cas~A region \citep{sibthorpe2010}.  However, that is not
necessary here, since we are exploiting the spatial correlations of dust
emission with color excess over the individual rectangles.  A corollary
is that we obtain the properties of the dust that is producing the
spatial variations within each rectangle; a uniform distribution would
not be detectable.

\subsection{Dust Temperature from the Diffuse Emission} \label{sedblast}

Small scale structures are remarkably well correlated across the three
BLAST bands. For a sufficiently large and homogeneous region an
estimate of the characteristic temperature can be obtained via
pixel-by-pixel correlations of images with respect to some reference
image (here BLAST 250~\micron).  We used the \emph{IDL} routine
\emph{SIXLIN} to perform the regressions, and adopted the bisector
slope as our specific estimator \citep{Isobe1990}.  The slopes of
these correlations describe the relative SED of the emission that is
changing in common in these images.

The SED for cold dust emission at temperature $\sim$ 15~K and
emissivity index $\beta = 1.8$ peaks at 200~\micron, and so in
principle the dust temperature could be determined from the curvature
in the SED through the three submillimeter passbands of BLAST.
Nevertheless, in practice it is always preferable to have broader
wavelength coverage, particularly on the short wavelength side of the
peak.  
At the shorter wavelengths we have examined the correlation slope of
the 100~\micron\ image from \IRAS\ with BLAST.  We used the reprocessed
IRIS product \citep{mairis}.  Because there was a hot point source in
rectangle 1 and another in rectangle 4 we used a version of the IRIS
images from which the sources have been removed (these hot sources are
not prominent at the BLAST passbands).
The IRIS 100~\micron\ image has $1\farcm5$ pixels in common with the
other maps used but a lower resolution of about 4\arcmin\
\citep{mamd2002}.  We considered using another \IRAS\ product, the
higher resolution HIRES image \citep{Cao1997}.  HIRES was originally
designed for resolving crowded compact sources \citep{Aumann1990} and
although empirically it improves the resolution of diffuse emission
too (e.g., \citealp{Martin1994}), this has not been quantified at the
few percent level.  We have, however, convolved the HIRES map
(resolution 2\arcmin) to the IRIS resolution and correlated it with
the original within the rectangles examined.  The slope was typically
0.96.  We concluded that using IRIS might result in a systematic error
of a few percent, but that this was acceptable given the much larger
calibration error of 13\% \citep{mairis}.  The effect of such
systematic errors is explored in Section~\ref{errors}.

\begin{figure}[t]
\includegraphics[width=\linewidth]{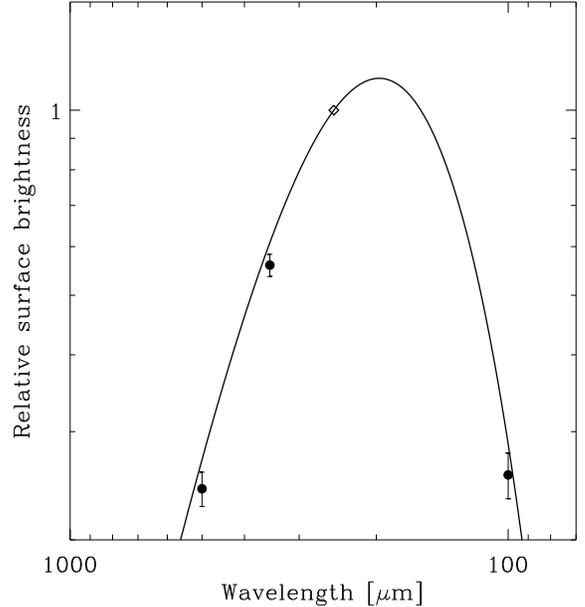} 
\caption{Relative SED from correlations in rectangle \rectno\ of BLAST
  and IRIS data with BLAST 250~\micron.  Implicitly the value is unity
  at 250~\micron\ (indicated by a diamond); there is no datum there
  used explicitly in the fit of the \emph{relative} SED.  With $\beta
  = 1.8$ the best fit temperature is \Tone. }
\label{sednorm}
\end{figure}

An example of the relative SED is shown in Figure~\ref{sednorm}, for
rectangle~\rectno.  Note that the fitting function implicitly passes
through unity at 250~\micron\ (indicated by a diamond); there is no
datum there used explicitly in the fit of the \emph{relative} SED.

The diffuse emission in the submillimeter and mid-infrared wavelengths
comes from two different dust components, distinguished principally by
their size distribution, namely Big Grains (BGs) and Very Small Grains
(VSGs) (\citealp{desert1990,lidraine2001,compiegne2011}).  The BGs, in
thermal equilibrium with the ambient radiation field, account for most
of the dust mass and therefore most of the longer wavelength
emission. VSGs are small enough to experience non-equilibrium heating
and so broaden the spectrum toward shorter wavelengths, beyond the
spectral peak of the BG emission.  VSGs comprise a relatively low
fraction of the total dust mass, even less in dense regions, and
typically their excess emission over what is expected from equilibrium
BGs alone appears at 60~\micron\ and shorter wavelengths; in the fields
studied here this emission is very faint.  Therefore, in this
exploratory work we adopted a single temperature SED and fit only data
for wavelengths 100~\micron\ and longer, using the \emph{IDL} routine
\emph{MPFIT} \citep{mark09}.  Even with fixed $\beta = 1.8$, this
grey-body functional form provides an acceptable fit to the data (e.g.,
Figure~\ref{sednorm}; the best-fit equilibrium temperature for this
rectangle is \Tone).

The observed low slope of the correlation of the 100~\micron\ IRIS image
with the 250~\micron\ BLAST image constrains the dust temperature to be
low.  The degree of correlation between the 100~\micron\ and the
250~\micron\ emission is somewhat less than it is between different
BLAST bands.  This decorrelation is probably due to a range of grain
temperatures within the volume sampled, possibly including a
contribution from non-equilibrium emission; temperature changes have
non-linear effects in the Wien tail.  The correspondingly larger
uncertainty in the slope of the correlation, by about a factor two,
means that the 100~\micron\ datum has less weight in fitting the SED.

\section{Observations of Color Excess}\label{extinction}

To determine the dust opacity from the submillimeter emission we need an
independent tracer of column density, here the near-infrared color
excess.  This can be estimated using $J$, $H$, and $K_{\rm s}$ data from
the 2MASS point source catalog, through a variety of techniques.
The color excess map used here was derived by the ``AvMAP'' procedure
which calculates the average near-infrared reddening of stars with a
method adapted from earlier analyses
\citep{lada1994,lombardi2001,cambresy2002}, with improvements as
described briefly in \citet{sch2006} and more completely in
\citet{Schneider2011}.  For diagnostic purposes, we have used this
procedure to produce maps of both \Ejh\ and \Ehk.  These maps have
resolution of about $2\farcm1$ and $1\farcm5$ pixels.

We have checked these maps against those more recently published for the
whole sky.  They correlate well with the extinction maps produced by
\cite{rowles2009} using a median near-infrared color excess technique
and with the color excess maps produced by \cite{dobashi2011} using the
``X'' percentile method.  For example, our \Ehk\ compared to that of
\cite{dobashi2011} is $1.08 \pm 0.01$ for the moderate column densities
of interest in our study (corresponding to \av~$< 10$), but with
considerable deviation beyond that, where extinction is harder to
determine.\footnote{Dobashi also combined his two color excess maps into
  a representation of $A_V$.  For the Vela region at least, we found
  that this is not well correlated with the underlying color excess
  maps.}

Our two color excess maps \Ejh\ and \Ehk\ are very tightly correlated,
with a slope of $1.77 \pm 0.01$
over a range up to \Ejk\ of 5.  The rectangles examined here had
maximum \Ejk\ typically less than a fifth of this value
(Table~\ref{tab2}).  Even over the whole field, there is no curvature
in the correlation as might be diagnostic of a medium where dense
clumps below the resolution limit of ``AvMAP'' systematically dimmed
stars to beyond the completeness limit at $J$ first, biasing \Ejh\ to
be low.

Theoretically, the \Ejh/\Ehk\ ratio is sensitive to the adopted shape
of the near-infrared extinction curve, often taken to be a power-law
in wavelength \citep{ccm1989,martinwhittet1990}, and to the filter
bandpasses and the intrinsic spectra of the stars being measured,
because of color corrections which change with increasing extinction.
Simulating all of these effects, we find that to explain the observed
color excess ratio, the power-law index for near-infrared extinction
is
$1.85 \pm 0.10$, encouragingly close to that found to be common from
studies of individual reddened stars \citep{martinwhittet1990}.
\cite{he1995} found a ratio of $1.64\pm 0.26$\footnote{Photometry was
  in the SAAO system but the ratio of the color excesses should be
  similar after color transformations to the 2MASS system; see summary
  by J. M. Carpenter at
  www.astro.caltech.edu/~jmc/2mass/v3/transformations/} for highly
obscured OB stars with \Ejk\ up to about 0.7.
\cite{indebetouw2005} obtained a value of $1.73\pm 0.2$ using various
analyses of 2MASS data probing to the much greater column densities
accessible in the infrared (\Ejk\ up to about 2.5).
The region that we have studied is therefore fairly normal, reinforced
by the \emph{relatively} moderate column densities in the rectangles chosen
(Table~\ref{tab2}).

Often near-infrared color excess maps are presented in terms of the more
familiar measure of extinction \av, via a ``total to selective
extinction'' conversion like
\begin{eqnarray}
A_V = R_{XY} E(X-Y).
\label{color}
\end{eqnarray}
From studies of individual stars sampling the local diffuse interstellar
medium the scaling coefficients frequently adopted, from
\citet{rieke1985} ignoring the slightly different filter sets, are
$R_{HK} = 15.87$, $R_{JH} = 9.35$, and $R_{JK} =
5.89$.\footnote{From these scaling coefficients, the ratio of \Ejh\ to
  \Ehk\ is 1.7, and the associated power-law index is 1.61
  \citep{ccm1989}.}
Even if the shape of the near-infrared extinction curve were fairly
universal, there are significant changes in the shape of the visual to
ultraviolet extinction curve, often parameterized in terms of $R_V
\equiv R_{BV}$, the ratio of total to selective extinction
\citep{ccm1989}.  Increasing $R_V$ from 3.1, characteristic of the
diffuse interstellar medium, to a value of 5.5, that might be more
typical of dark clouds, lowers the scaling coefficients by 20\%.
Furthermore, there are color corrections at high column densities.  We
have no direct evidence what scaling to \av\ would be valid for these
Galactic lines of sight toward Vela.  Therefore, we decided to use the
sum of our two maps, \Ejk, as the best measure of column density,
staying close to the directly observable color excesses, mitigating any
hidden effects of grain evolution, and more generally avoiding
unnecessary, often hidden, assumptions.  Of course, \Ejk\ still has to
be calibrated to give $N_{\rm H}$ (Section~\ref{sec:hydrogen}).

As summarized in Table~\ref{tab1}, within the various rectangles
measured, \Ejk\ ranges typically from 0.24 to 1.1~mag, reaching as low
as 0.17~mag in the most diffuse field (rectangle 11) and as high as
3.7~mag toward the main dense cloud in Vela~C (rectangle B).  
In a more familiar metric, these color excesses would correspond to
\av\ = 0.94, 1.4, 6.5, and 22~mag, using the above-mentioned scaling
coefficients.
However, such scaling needs to regarded with caution, because the high
values of color excess and extinction are much beyond those toward stars
for which the full infrared to ultraviolet extinction curve, the
underlying dust size distribution, and the relationship of \av\ to color
excesses and $N_{\rm H}$ have been studied directly.


\section{Emission and Color Excess}\label{correlation}

\begin{figure}[t]
\includegraphics[width=\linewidth]{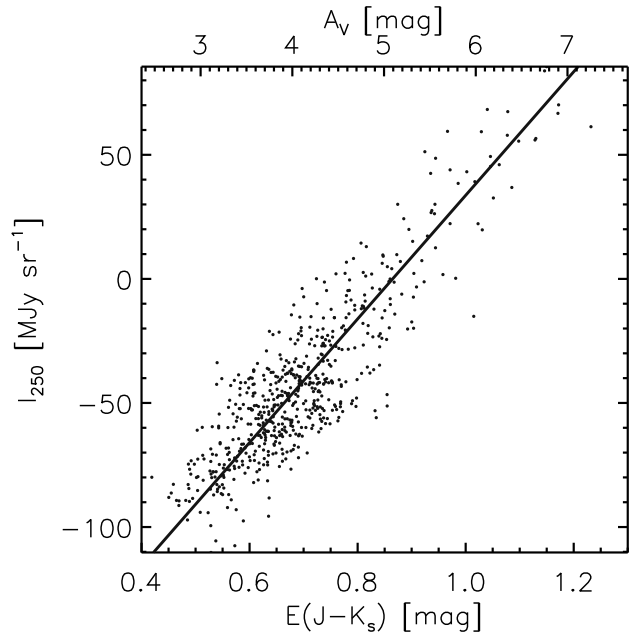} 
\caption{Pixel-pixel correlation of 250~\micron\ BLAST emission with the
  color excess \Ejk, in rectangle \rectno.  The solid line has a best-fit
  slope $\epsilon_{\rm c}(1200) = $ \epsilonc~MJy~sr$^{-1}$~mag$^{-1}$.
The $A_V$ scale along the top is only approximate, but is provided for
context.  }
\label{dustvsAv}
\end{figure}

Figure~\ref{dustvsAv} illustrates the correlation of
$I_{250}$,\footnote{Actually $I_\nu$ but labelled with the wavelength
  of the passband.} the continuum emission at 250~\micron, with the
\Ejk\ color excess.  Recalling that the dust emission probes the whole
line of sight, while the depth to which color excess probes is in
principle limited by both attenuation and sensitivity, the correlation
is remarkable.  The slope of the correlation characterizes the
emissivity $\epsilon_{\rm c}(1200)$ of the dust that is causing the
spatial variations in column density across the map.  The zero point
of the BLAST maps is not needed, but by the same token any fairly
uniform screen of material cannot be characterized by this correlation
technique.

The dispersion about the fit in Figure~\ref{dustvsAv} is 0.06~mag if
measured horizontally or 16~MJy~sr$^{-1}$ vertically.
We have estimated the error in the \Ejk\ map using the scatter about
the correlation between the two maps \Ejh\ and \Ehk, examined
rectangle by rectangle.  From this approach, the rms in the \Ejk\ map
is $\sim$ 0.03 mag.  Possibly this is an underestimate because of
correlated errors.  Therefore, as an alternative we looked at the rms
in the adopted \Ejk\ map in relatively smooth regions of low color
excess and found 0.05 mag; this could overestimate the error because a
part is due to real cirrus fluctuations.  For comparison with the rms
for $I_{250}$ below, 0.03 -- 0.05~mag corresponds to 7 --
12~MJy~sr$^{-1}$.
Using similar strategies we have estimated the error in $I_{250}$:
from the scatter about the correlation of the $I_{250}$ map with the
$I_{350}$ map, rectangle by rectangle, we estimate the rms in
$I_{250}$ to be 2~MJy~sr$^{-1}$; and the rms fluctuation in the
dimmest part of the $I_{250}$ map is 3~MJy~sr$^{-1}$.
Returning to Figure~\ref{dustvsAv}, our estimates suggest that the
error in \Ejk\ makes the larger contribution of the two and that taken
together there is no compelling reason to require some additional
cosmic scatter about the correlation.

The correlations with the other BLAST bands are equally good, but, as
foreshadowed by the above cross-band comparisons
(Section~\ref{sedblast}), the degree of correlation of 100~\micron\
emission with \Ejk\ is less, again possibly due to changes in
temperature within the volume probed.

Even though we chose rectangles to avoid strong sources, we explored
the possibility that clumpy dust that would contribute to the BLAST
emission map might be missed in the sampling of stars underlying the
\Ejk\ map.  We found no evidence for ``excess'' emission at high color
excess even in rectangle B.  In the latter rectangle there is possibly
a slight deficit, related to clumps of cool dust; in that extreme case
our assumption of a constant temperature for dust within the
rectangle, rather than systematics in estimating the color excess, is
probably the limiting approximation.

\subsection{SED}\label{sed2mass}

\begin{figure}[t]
\includegraphics[width=\linewidth]{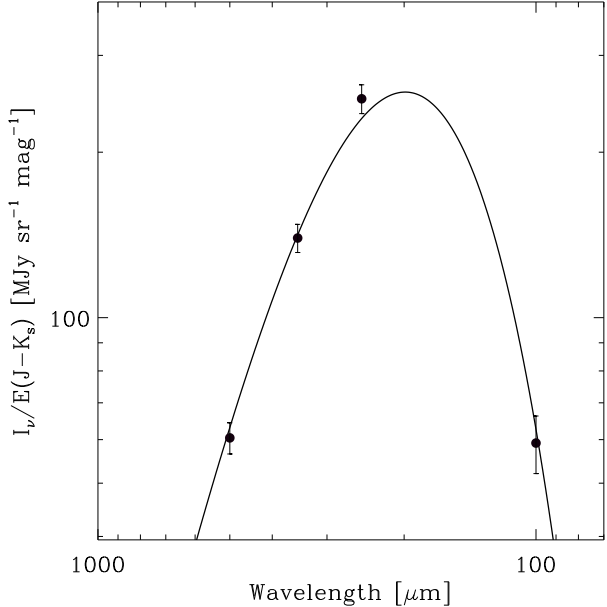} 
\caption{SED from the slopes of the correlations of dust emission with
  near-infrared color excess \Ejk\ for rectangle~\rectno.  The
  best-fit temperature, \Tonebav~K, is similar to that from the
  cross-band correlations (Figure~\ref{sednorm}).  The SED amplitude
  $\epsilon_{\rm c}(1200)$ = \epsiloncamp\ MJy~sr$^{-1}$~mag$^{-1}$ is
  close to the value obtained from direct correlation
  (Figure~\ref{dustvsAv}).  }
\label{sedmodel}
\end{figure}

Fitting the slopes of the correlations between dust emission and color
excess to a parameterized single-temperature SED yields the dust
temperature recorded in Table~\ref{tab2}.  An example is given in
Figure~\ref{sedmodel}.  These temperatures from the SED
fits are around 15~K, definitely cooler than the 17.9~K typical of the
high latitude diffuse interstellar medium \citep{abergelDd2011}. In
Table~\ref{tab2} we also present the emissivity $\epsilon_{\rm c}(1200)
= I_{250}/E(J-K_{\rm s})$, the best-fit amplitudes obtained from the
SEDs.  Ultimately (Section~\ref{results}) we will derive the opacity
from these observables.

As a consistency check we verified that $T$ derived from the
$I_\nu$ -- \Ejk\ correlations here is close to that obtained via the
relative SED from the cross-band emission-map correlations
(Section~\ref{sedblast}).  This supports our premise that relative
changes in column density are equally well sampled by changes in BLAST
and IRIS emission and in near-infrared excess.

\begin{deluxetable*}{cccccccccc}
\tablewidth{0pt}
\setlength{\tabcolsep}{0.01in}
\tablecaption{Dust emission in the Galactic plane toward Vela}
\tablehead{
\colhead{ID}&
\colhead{$T$}&
\colhead{$\epsilon_{\rm c}(1200)$\tablenotemark{$\rm a$}}&
\colhead{$r\kappa_{0}$}&
\colhead{$\sigma_{\rm e}(1200)$}&
\colhead{$P$}&
\colhead{$E(J-K_{\rm s})$ range}\\
\colhead{}&
\colhead{K}&
\colhead{MJy~sr$^{-1}$~mag$^{-1}$}&
\colhead{cm$^{2}$~gm$^{-1}$}&
\colhead{10$^{-25}$~cm$^2$~H$^{-1}$}&
\colhead{$10^{-31}$~W~H$^{-1}$}&
\colhead{mag}\
}
\startdata
       1&16.8$\pm$ 0.4&202.6$\pm$  7.6&0.09$\pm$0.01& 2.0$\pm$0.2& 5.3$\pm$ 0.4&0.25/0.68\\
       2&15.5$\pm$ 0.2&252.1$\pm$  6.6&0.15$\pm$0.01& 3.4$\pm$0.2& 5.6$\pm$ 0.2&0.34/0.85\\
       3&15.2$\pm$ 0.5&190.1$\pm$ 11.0&0.12$\pm$0.01& 2.8$\pm$0.3& 4.0$\pm$ 0.4&0.17/0.59\\
       4&16.2$\pm$ 0.2&265.8$\pm$  6.1&0.13$\pm$0.01& 3.0$\pm$0.1& 6.5$\pm$ 0.2&0.24/0.71\\
       5&13.8$\pm$ 0.7& 83.8$\pm$  8.6&0.08$\pm$0.01& 1.8$\pm$0.3& 1.5$\pm$ 0.3&0.31/0.59\\
       6&14.9$\pm$ 0.2&164.5$\pm$  4.1&0.11$\pm$0.01& 2.6$\pm$0.1& 3.4$\pm$ 0.1&0.34/0.85\\
       7&15.2$\pm$ 0.2&230.3$\pm$  7.2&0.14$\pm$0.01& 3.3$\pm$0.2& 4.9$\pm$ 0.3&0.51/1.10\\
       8&15.1$\pm$ 0.2&181.6$\pm$  5.4&0.12$\pm$0.01& 2.7$\pm$0.2& 3.8$\pm$ 0.2&0.42/0.85\\
       9&15.1$\pm$ 0.2&249.0$\pm$  7.2&0.16$\pm$0.01& 3.7$\pm$0.2& 5.3$\pm$ 0.2&0.25/0.59\\
      10&15.2$\pm$ 0.4&124.7$\pm$  6.6&0.08$\pm$0.01& 1.8$\pm$0.2& 2.6$\pm$ 0.2&0.34/0.68\\
      11&15.2$\pm$ 0.1&263.6$\pm$  5.0&0.16$\pm$0.01& 3.9$\pm$0.1& 5.6$\pm$ 0.2&0.17/0.42\\
      12&15.1$\pm$ 0.2&189.0$\pm$  4.8&0.12$\pm$0.01& 2.8$\pm$0.2& 4.0$\pm$ 0.2&0.34/0.81\\
 \hline\hline \\
       A&12.4$\pm$ 0.2& 75.3$\pm$  2.9&0.11$\pm$0.01& 2.6$\pm$0.2& 1.2$\pm$ 0.1&0.42/1.70\\
       B&12.2$\pm$ 0.3& 99.9$\pm$  5.3&0.16$\pm$0.02& 3.8$\pm$0.4& 1.5$\pm$ ~0.1&1.19/3.74\\

\enddata
\label{tab2}
\tablenotetext{a}{Best fit amplitude from fitting SED}
\end{deluxetable*}

\subsubsection{SED weighting and parameter errors}\label{correlatederrors}

The slopes of the correlations of BLAST and \IRAS\ with \Ejk, and
their uncertainties, were obtained using \emph{SIXLIN}.  For a given
rectangle the uncertainties were similar for all BLAST bands and about
half of that for the \IRAS\ band.  This ratio persisted for all
rectangles and so, as in Section~\ref{sedblast}, we have adopted this
ratio uniformly in performing weighted SED fits. We consider the
systematic effect of calibration errors between BLAST and \IRAS\ in
Section~\ref{errors}.  Here a reduced $\chi^2$ of about unity for the
fit typically requires an increase of the nominal \emph{SIXLIN}
uncertainties to about 6\% and 12\% for the BLAST and \IRAS\ bands,
respectively.  This increase results in more conservative errors on
the model parameters $T$ and $\epsilon_{\rm c}(1200)$ (ultimately
scaled to $\sigma_{\rm e}(1200)$) and on the dependent quantity $P$
(Equation~(\ref{power})), as calculated by Monte Carlo simulation
\citep{chapin2008}.  A set of 500 realizations of mock data is
generated starting with the actual slopes and adding Gaussian noise
with the above uncertainties.  For each realization the SED was fit
and the corresponding parameters recorded.  Finally, the uncertainty
on each quantity was obtained by fitting a Gaussian to the histogram
of the generated distribution. These are the statistical errors
reported in Table~\ref{tab2}.  By keeping a record of each fit we also
tracked the correlations of the errors and so can produce the
elliptical 1-$\sigma$ confidence intervals in, for example, the $T$ --
$\sigma_{\rm e}(1200)$ plane (see Section~\ref{relation}).


\section{Hydrogen Column Density and Color Excess}\label{sec:hydrogen}

To calculate the opacity, $\sigma_{\rm e}(\nu)$, we need to relate the
color excess \Ejk\ to the column density of hydrogen, $N_{\rm {H}}$.
Often this is obtained by converting \Ejk\ to \av\ and then using the
ratio $N_{\rm {H}}/A_V$, all for values found in the local diffuse
interstellar medium.  This is arguably not justified for the high column
density lines of sight where it is applied, and therefore is worthy of
some reflection.

$N_{\rm {H}}$ toward individual stars has been measured using
Lyman~$\alpha$ absorption and the absorption lines of molecular hydrogen
using \emph{Copernicus} and the \emph{Far Ultraviolet Spectroscopic
  Explorer} (\emph{FUSE})
\citep{bohlin1978,savage1977,diplas1994,rachford2002,rachford2009}.  But
sensitivity requirements for these ultraviolet measurements have limited
the column density probed to \av\ about 3~mag or \Ejk\ about 0.5~mag,
lower than in the fields studied here, despite these fields being
selected for relatively low column density within the Vela map.  If the
higher column density here were simply the result of a long line of
sight through diffuse material, then it might be argued that the
material would be similar to what has been probed directly using
individual stars.  However, the lines of sight within the rectangles are
at least partially molecular \citep{yamaguchi1999}, and so the material
is probably more localized along the line of sight with higher spatial
density.

Relating $N_{\rm {H}}$ to \Ebv, these studies of individual stars have
found that
\begin{equation}
N_{\rm {H}} = 5.8 \times 10^{21} E(B-V)\, {\rm cm^{-2}}.
\label{htotvsebv}
\end{equation}
Equation~(\ref{htotvsebv}) can be recast in terms of \av\ if $R_V$
is known for individual lines of sight, but we think that this is even
less appropriate for our application.  Evidence for dust evolution on
lines of sight passing through dense material comes from changes in the
optical-ultraviolet extinction curve (\citealp{kimmartin1996} and
references therein), which can be parameterized by the changes in
$R_V$ \citep{ccm1989}.  Because of this, one might expect
deviations from a simple linear relationship between $N_{\rm {H}}$ and
\av\ or even \Ebv, especially for dense regions with high column
density.  There is an indication of an increase in the slope in the
Equation~(\ref{htotvsebv}) correlation to $6.6 \times
10^{21}$~cm$^{-2}$~mag$^{-1}$ for some higher column density lines of
sight (\av\ of 1 to 5 mag) studied with \emph{FUSE}
\citep{rachford2002,rachford2009}.  However, it is not known whether
this trend is maintained in more dense regions where probing the total
hydrogen column density directly is not possible.

Because the ratios of near-infrared color excess to $N_{\rm {H}}$
probably change less significantly as the grains evolve
\citep{martinwhittet1990, kimmartin1996}, it seems advantageous to
examine directly the correlation between $N_{\rm {H}}$ and our
observable \Ejk.

We used atomic as well as molecular hydrogen column densities measured
by \citet{savage1977} and \cite{rachford2002,rachford2009}.  Below a
threshold of $N_{\rm {H}} = 0.6\times 10^{21}$~cm$^{-2}$, as noted by
\cite{savage1977}, the majority of hydrogen is in atomic form.  For
higher column densities, with a mixture of conditions along the line
of sight, our best-fit approximation to the trend in the data above
the threshold where both forms of hydrogen are measured is
\begin{equation}
N_{\rm H}=1.52 \times (N_{\rm HI} - 0.6 \times 10^{21}) + 0.6 \times
10^{21} \,  {\rm cm^{-2}}.
\label{diplas}
\end{equation}
\cite{diplas1994} measured atomic hydrogen along many more lines of
sight for which \Ejk\ can be obtained, but not molecular hydrogen, and
so Equation~(\ref{diplas}) was used to make a correction. At any
column density above the threshold there is dispersion in the
fractional amount of hydrogen in molecular form and so this correction
is valid only statistically.

For the program stars of \cite{diplas1994}, we extracted archival
2MASS photometric measurements via
\emph{GATOR}\footnote{http://irsa.ipac.caltech.edu/applications/Gator/}.
We concentrated on the $J$ and $K$ bands to maximize the differential
extinction and to take advantage of the somewhat better photometry
than for the $H$ band.  There is a bimodal distribution in the
reported photometric uncertainty, the higher peak relating to
saturation for the brighter O and B stars.  Based on the lower peaks
at $J$ and $K$, characterizing the normal photometric error, we have
selected those sources which have a combined error in $J - K$ less
than $3\sigma_{JK}$, i.e., 0.025~mag.  To find the color excess \Ejk\
we used the dependence of intrinsic colors on spectral classification
given by \citet{straizys2009}.

A preliminary plot of \Ebv\ vs. \Ejk\ showed that some of these
selected program stars are much redder in \Ejk\ than could be expected
from interstellar extinction.  We confirmed from the spectral
classification that most of the anomalous stars are known Be or
emission-line stars; these have near-infrared emission in addition to
that from the photosphere.  Hence, to refine our source selection we
excluded sources in the color excess plane lying beyond $2.5 \sigma$
from the correlation line $E(B-V)/E(J-K_{\rm s}) = 1.9$; this
precaution serves to exclude all of the Be stars from our final list,
without unnecessarily biasing our results below.

\begin{figure}
\includegraphics[width=\linewidth]{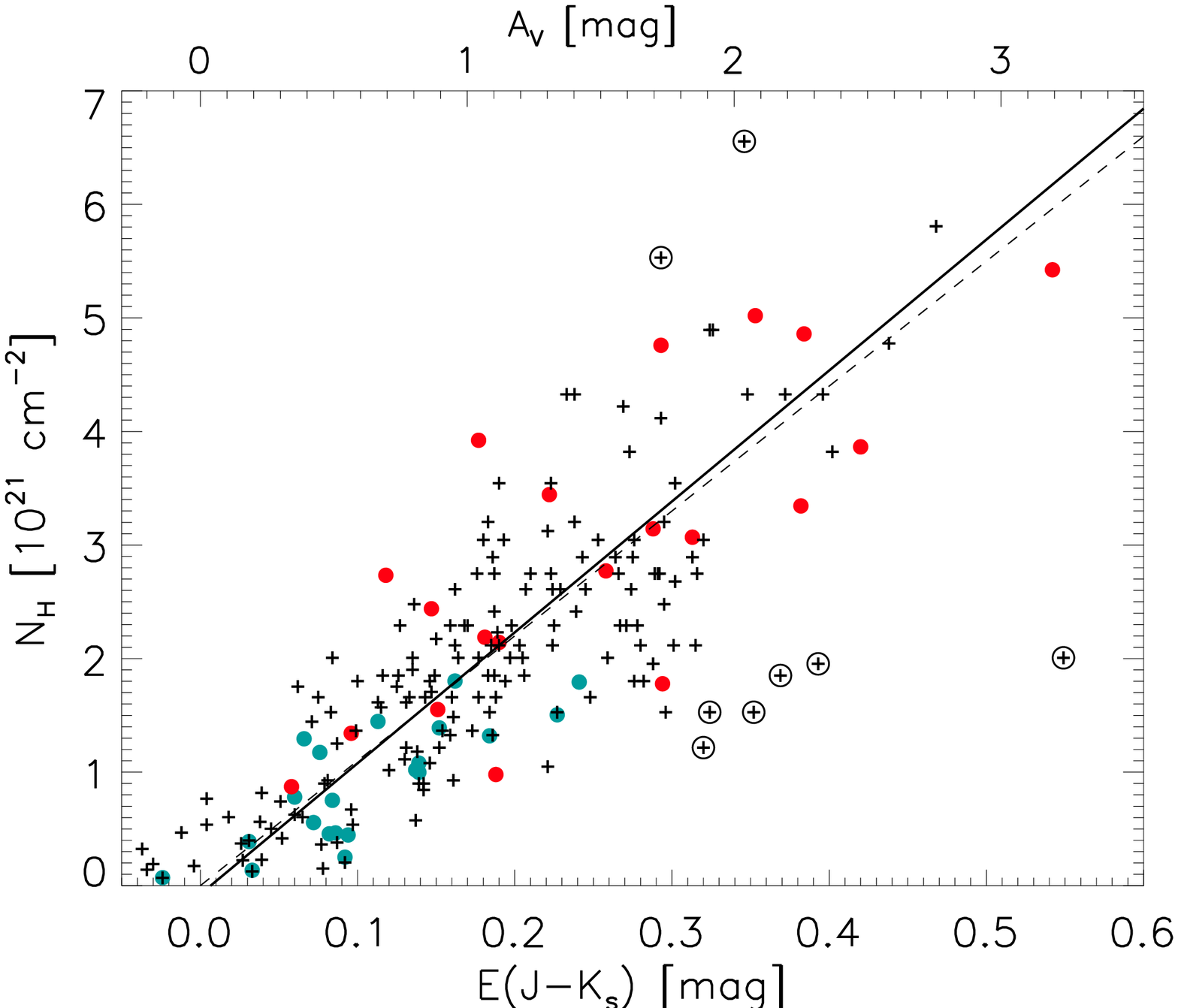} 
\caption{Correlation between $N_{\rm {H}}$ and \Ejk.  
Solid green and red circles represent $N_{\rm {H}}$ measurements from
\cite{savage1977} and \cite{rachford2002,rachford2009}, respectively.
Plus symbols are for $N_{\rm {H}}$ from atomic hydrogen measurements by
\cite{diplas1994}, corrected statistically using Equation~(\ref{diplas}).
The best-fit line (solid) has slope ($11.5 \pm 0.5) \times
10^{21}$~cm$^{-2}$~mag$^{-1}$.  The dashed line with slope $11.0 \times
10^{21}$~cm$^{-2}$~mag$^{-1}$ is from Equation~(\ref{htotvsebv}) converted
using standard color ratios (see text).  Open circles mark the data
excluded in the process of iterative fitting. 
The $A_V$ scale along the top is only approximate, but illustrates the
limited range covered in this calibration compared to that found in the
color excess map in Figure~\ref{fig:extinction}, even within the
rectangles analyzed.  }
\label{fig:nhvsejk}
\end{figure}

Figure~\ref{fig:nhvsejk} shows a good correlation between $N_{\rm {H}}$
and \Ejk. We have used an iterative fitting technique to identify and
exclude a few outliers more than $3\sigma$ from the correlation. Not
surprisingly, and justifying this approach, the outliers are stars from
\cite{diplas1994}, where we have estimated the total hydrogen column
density from the atomic hydrogen column density measurements using
Equation~(\ref{diplas}).

The best-fit line is
\begin{equation}
\left(\frac{N_{\rm {H}}}{10^{21}\,  {\rm cm^{-2}}}\right) = (11.5 \pm 0.5) E(J-K_{\rm s})  -0.07 \pm 0.01.
\label{eq:nhvsejk}
\end{equation}
For comparison, conversion of the slope in Equation~(\ref{htotvsebv})
using standard color ratios appropriate to an interstellar extinction
curve with $R_V = 3.1$ \citep{rieke1985,ccm1989} gives a slope $11.0
\times 10^{21}$~cm$^{-2}$~mag$^{-1}$.  This is close to what we have
found directly, perhaps reassuringly, but also to some degree
coincidentally, given the range of conditions sampled (the converted
value would be only half as large for $R_V = 5.5$).  Note the
significant dispersion about the average relation.  

However, it has only been possible to obtain a direct calibration of
this relation to \Ejk\ about 0.4 and so applications at higher column
densities need to be viewed with caution. This includes the present
application, where the \emph{top} of the \Ejk\ range in most
rectangles (see Table~\ref{tab2}) is beyond the calibrated range.  On
the one hand, to the extent that the larger values are simply the
result of longer pathlengths, the calibration should stand. But, on
the other hand, if the larger values of \Ejk\ are the result of
increased volume density then the grains might evolve.  For example,
calculations of time-dependent extinction curves resulting from grain
evolution by ice-mantle formation and aggregation \citep{ormel2011}
show how \Ejk\ might increase for a given column of material, at least
initially, which would decrease the ratio $N_{\rm {H}}$/\Ejk.  This
would in turn raise the derived opacity (Equation~(\ref{final})),
exaggerating the changes reported below.

The same caution about lack of direct calibration also holds for any
application which uses such measures of infrared color excess to gauge
column density.  The situation is further muddied, unnecessarily, when
the column density is cited in terms of $A_V$.
 

\section{Results}\label{results}

\subsection{Opacity at  1200~GHz or 250~\micron}\label{opacity}

Using the temperatures and the amplitudes from the SED fits in
Section~\ref{correlation}, together with the ratio $N_{\rm {H}}/E(J-K_{\rm
  s})$ from Section~\ref{sec:hydrogen}, we calculated the opacity from
\begin{equation}
 \sigma_{\rm e}(1200) = \epsilon_{\rm c}(1200)/[B_{\nu_0}(T) \times N_{\rm
     {H}}/E(J-K_{\rm s})].
\label{final}
\end{equation}
Recall that $\sigma_{\rm e}(1200) \equiv \mu m_{\rm H} r \kappa_0$.  The
derived values are recorded in Table~\ref{tab2} along with their Monte
Carlo errors (Section~\ref{correlatederrors}).  The typical
opacity in these regions is about $2.8 \times 10^{-25}$~cm$^2$~H$^{-1}$
or equivalently 0.12~cm$^2$~gm$^{-1}$.  There are considerable
variations above what can be accounted for by the errors.  Furthermore,
all values are significantly above what is typical of the high latitude
diffuse interstellar medium, $1.0 \times 10^{-25}$~cm$^2$~H$^{-1}$
\citep{abergelDd2011}.

Rectangle A coincides with cloud 28 of \cite{yamaguchi1999} for which
\cite{netterfield2009} have estimated the dust opacity to be $r \kappa_0
= 0.16$~cm$^2$~gm$^{-1}$ by comparing the integrated submillimeter dust
emission with the total mass of gas estimated from the CO emission.  The
latter introduces an uncertainty of a factor 1.5 -- 2.  Our new value is
$0.11 \pm 0.01$~cm$^2$~gm$^{-1}$.

\subsection{Integrated emission}\label{luminosity}

A physical quantity of interest is the total energy emitted by dust per
hydrogen atom, $P$ (Equation~(\ref{power})).  For the diffuse high
Galactic latitude interstellar medium the value is fairly uniform near $3.8
\times 10^{-31}$~W~H$^{-1}$ \citep{abergelDd2011}.  The typical value
found here (see Table~\ref{tab2})
is somewhat higher, $4.5 \times 10^{-31}$~W~H$^{-1}$.  However, not
surprisingly, there is considerable variation in the Galactic plane (see
also Section~\ref{pandopacity} and Figure~\ref{fig:usigma} below).

\subsection{Relationships}\label{relation}

The parameters $T$, $\sigma_{\rm e}(1200)$, and $P$ for any rectangle
are related at a fundamental level through Equation~(\ref{power}).
Because we have used a fixed $\beta = 1.8$, this relationship can be
quantified as
\begin{equation}
(T/T_0)^{5.8} = (P/P_0)\, (\sigma_{\rm e}(1200)/\sigma_0)^{-1},
\label{locus}
\end{equation}
using the above-mentioned high latitude diffuse ISM values for
normalization (Section~\ref{dustemission}).

\begin{figure}
\includegraphics[width=\linewidth]{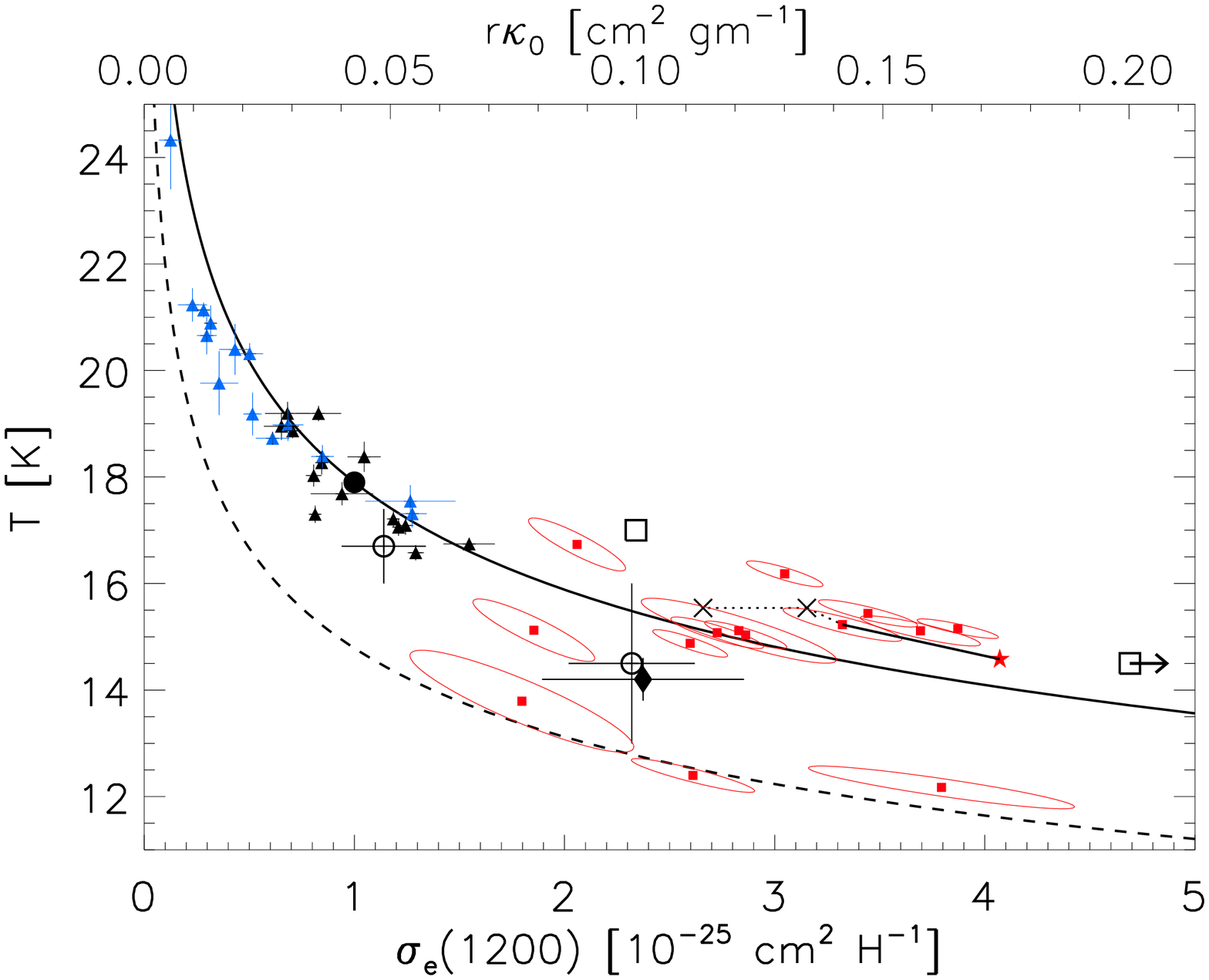} 
\caption{Dust temperature, $T$, vs.\ emission opacity, $\sigma_{\rm e}(1200)$.
Values for the regions studied here in the Galactic plane toward Vela
are represented by the red filled squares with elliptical 1-$\sigma$
confidence intervals.
Loci are for constant $P$, the third parameter (Equation~(\ref{locus})):
solid, $3.8 \times 10^{-31}$~W~H$^{-1}$ for the diffuse high latitude
ISM \citep{abergelDd2011}; and dashed, three times less.
If $\beta$ were fixed at 2.0 in the SED fit, each solution shifts
systematically at roughly constant $P$; this is illustrated
for rectangle~\rectno\ by a line connecting to the star.
Also for rectangle~\rectno, the systematic effect of a 15\%
calibration error between BLAST and IRIS is illustrated by the dotted
line segments (see text).
The filled circle locates the standard values for the diffuse high
latitude ISM \citep{abergelDd2011}.
Values for individual diffuse high latitude regions \citep{abergelDd2011}
are also plotted, with black and blue distinguishing between LVC and IVC
components, respectively.
Some further comparisons with empirical results
(Appendix~\ref{literature}) are plotted.  Open circles show
\emph{Planck} results from \cite{abergelTd2011} toward the Taurus
molecular cloud for lines of sight both atomic (upper left) and
molecular; the vertical error bars indicate the ranges of temperature.
The filled diamond is another estimate for the Taurus region using
\emph{Spitzer} \citep{terebey2009}.
The squares show the range of values found using \emph{Herschel} data
in the environs of a \emph{Planck} cold clump \citep{juvela2011};
these results support the trend with $T$ (and column density)
established by the other data and extend it to even higher
$\sigma_{\rm e}(1200)$ at the highest column densities.
}
\label{fig:tsigma}
\end{figure}

\begin{figure}
\includegraphics[width=\linewidth]{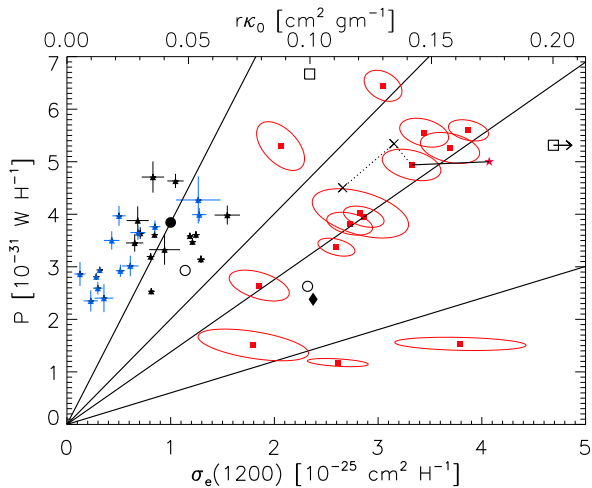} 
\caption{Like Figure~\ref{fig:tsigma} but for power emitted by dust
  per hydrogen atom, $P$, vs.\ emission opacity, $\sigma_{\rm
    e}(1200)$, and with loci (Equation~(\ref{locus})) for different $T$:
  13, 15, 16, and 17.9~K from lower to upper.  }
\label{fig:usigma}
\end{figure}

The parameters derived for the different rectangles (Table~\ref{tab2})
and their elliptical 1-$\sigma$ confidence intervals
(Section~\ref{correlatederrors}) are displayed in two complementary
diagrams, $T$ -- $\sigma_{\rm e}(1200)$ (Figure~\ref{fig:tsigma}) and
$P$ -- $\sigma_{\rm e}(1200)$ (Figure~\ref{fig:usigma}).  These
figures include loci according to Equation~(\ref{locus}) along which
the third parameter is constant.
Because of fixed $\beta$ and Equation~(\ref{locus}), the third possible
diagram, $T$ -- $P$, contains no independent new information. 

\subsection{Other errors}\label{errors}

Throughout this study we have avoided high extinction regions, owing
to a larger uncertainty in the color excess and in its conversion to
$N_{\rm H}$.

The adopted ratio of $N_{\rm H}/E(J-K_{\rm s})$ might not be
universally applicable.  A systematic 5\% uncertainty in this ratio,
or a larger error for individual measurements given the dispersion
about the relation (Figure~\ref{fig:nhvsejk}; also relevant to
Equation~(\ref{htotvsebv}) which is just the average), induces
(inversely) errors of the same size in $\sigma_{\rm e}(1200)$. This is
potentially a major contribution to the error budget.  While we might
have minimized this uncertainty successfully by avoiding high
extinction regions, by the same token our derived values of opacity
might not be representative of the unanalyzed high extinction regions,
including compact sources.  Because of the effects of grain evolution
mentioned in Section~\ref{sec:hydrogen} (see also
Section~\ref{insight}), any application where near-infrared color
excess is used to estimate high column densities needs to be assessed
critically.
We have investigated whether part of the trends seen in
Figures~\ref{fig:tsigma} and \ref{fig:usigma} might be induced by
changes in $\sigma_{\rm e}(1200)$ originating in a trend of $N_{\rm
  H}/E(J-K_{\rm s})$ with environment; however, a scatter plot of
$\sigma_{\rm e}(1200)$ vs.\ the average $<E(J-K_{\rm s})>$ for the
different rectangles shows no definitive trend, as can be judged by
inspection of Table~\ref{tab2} as well.

We kept the rectangles fairly small so that the assumption of a uniform
temperature for the dust was at least plausible; however, this does not
control for temperature changes along the line of sight.

There are calibration errors which could produce systematic effects.
For the BLAST bands these are 8.1, 7.1, and 7.8\% at 250, 350, and
500~\micron, respectively, and by the nature of the calibration
technique they are well correlated between the bands
\citep{truch2009}.  The calibration error of the IRIS 100~\micron\
image used is 13\% \citep{mairis}.  Because the calibration techniques
are independent, there is the possibility of a systematic effect when
SEDs are fit to the combined data.
If IRIS were adjusted upward, then $T$ would be higher and $\sigma_{\rm
  e}(1200)$ lower.  If instead it were BLAST that was adjusted downward,
by the same relative amount, then the same higher $T$ would be found and
$\sigma_{\rm e}(1200)$ would be even lower.  Such a two-part trajectory
is illustrated in the figures for rectangle~\rectno, for a relative
calibration error lowering BLAST/IRIS by 15\%.

For consistency and comparison to other studies we adopted $\beta =
1.8$. The choice of $\beta$ affects the parameters derived from the
SED fit in a systematic way.  The total power radiated $P$ is quite
insensitive to changes in $\beta$, because the SED is required to fit
the data across the whole range being integrated.  However, both the
amplitude and temperature change, the latter affecting the derived
opacity more profoundly because of the non-linearity of the Planck
function (Equation~(\ref{final})).
For example, if $\beta$ were increased from 1.8 to 2.0, the
temperature would be systematically lower and the opacity
systematically higher.  This shift is illustrated in
Figure~\ref{fig:tsigma} for rectangle~\rectno.  Such effects seem
unlikely to account for the much larger values of the opacity found
here compared to those in the high latitude interstellar medium.
On the contrary, in Appendix~\ref{abeta} we explore the potential
effect of $\beta$ further, in particular through a possible
$\beta$ -- $T$ relationship, and conclude that this would exaggerate the
changes in opacity already found with fixed $\beta$.  From this
perspective, the estimated magnitude of the changes found and
discussed below is conservative.


\section{Discussion}\label{analysis}

There are significant variations from rectangle to rectangle in the derived
parameters $T$, $\sigma_{\rm e}(1200)$, and $P$.  

For comparison with these new results for the Galactic plane toward
Vela, the filled circle in Figures~\ref{fig:tsigma} and
\ref{fig:usigma} locates the standard values for the diffuse high
Galactic latitude ISM from \cite{abergelDd2011} that were used to
normalize Equation~(\ref{locus}).  The data plotted for individual
diffuse high latitude regions are from SED fits ($\beta = 1.8$) to the
emissivities reported in Table~2 of \cite{abergelDd2011}, in which
local velocity clouds (LVC) and intermediate velocity clouds (IVC)
have been separated.  Note that the dust emission underlying these
SEDs is measured from a combination of \emph{Planck} and IRIS data.
Even at high latitudes there is considerable variation, but
interestingly the ranges of $T$ and $\sigma_{\rm e}(1200)$ do not
overlap with those found in the present study.

Overall there is a trend in Figure~\ref{fig:tsigma} of decreasing $T$
with increasing $\sigma_{\rm e}(1200)$.  Although the error ellipses
are aligned roughly along this trend, the values span a much larger
range than can be attributed to the individual errors of the
independent measurements.
The range of $P$ in the present study extends to both higher and lower
values than seen at high latitudes.

Other estimates discussed in Appendix~\ref{literature} are plotted in
the figures.  These support the trend of decreasing $T$ with
increasing $\sigma_{\rm e}(1200)$ seen above.

It is useful to think of Equation~(\ref{locus}) in terms of
\emph{cause} (on the right hand side) and \emph{effect}, $T$.  The $P$
calculated here is for emission by big grains in thermal equilibrium
and is therefore also the total power absorbed by these grains when
exposed to the interstellar radiation field (ISRF).  The opacity
$\sigma_{\rm e}(1200)$ measures an intrinsic property of the big
grains, how efficiently they can emit. This emission opacity and the
absorbed $P$ that needs to be emitted determine what the equilibrium
$T$ must be.
Thus it can be seen that big grains will be cooler in a less intense
ISRF, and/or if they were to evolve to have a lower absorption opacity
and/or a larger emission opacity.

Here we examine the evidence for such changes using the complementary
diagnostic Figures~\ref{fig:tsigma} and \ref{fig:usigma}.
Considering $T$ as an \emph{effect}, we concentrate further discussion
on the causes, $\sigma_{\rm e}(1200)$ and $P$.

\subsection{Changes in $\sigma_{\rm e}(1200)$}\label{tandopacity}

A main goal of this paper was to quantify the dust emission opacity in a
new environment near the Galactic plane, that is of higher column
density than the high latitude ISM, and at least partly molecular.  We
have found that in this environment $\sigma_{\rm e}(1200)$ is higher by
typically a factor three and, extending the finding by
\cite{abergelDd2011}, it changes from region to region.

Changes in the emission opacity are certainly intriguing but not
understood.  One possibility is that the opacity is raised when grains
aggregate, changing the basic structure to something more porous and
fractal than homogeneous \citep{ossenkopf1994}.  This has been discussed
for dense molecular clouds, where grains also develop ice mantles
\citep{ormel2011}, but its relevance to such evolutionary changes
occurring even within the more diffuse medium seems less
obvious. Perhaps one needs a change in perspective on the direction of
grain evolution, regarding the higher values found here as ``normal''
for dense regions and the lower values as the result of evolution of
dust back toward a different state in the diffuse interstellar medium.
See \citet{Jones2009} for a related discussion on extinction curves.

\subsection{Changes in $P$}\label{pandopacity}

One of the possible reasons for the range of values of $P$ is
variation of the interstellar radiation field (ISRF) in the Galactic
plane.  Attenuation in dense molecular clouds seems the likely cause
of the lower $P$ in rectangles A and B.  Likewise, it is at least
plausible that the ISRF is higher in those regions (rectangles 2, 4,
9, 11) with $P$ significantly larger than $5 \times
10^{-31}$~W~H$^{-1}$. 
However, as can be seen from Table~\ref{tab2}, a scatter plot of $P$
vs.\ the average $<E(J-K_{\rm s})>$ for the different rectangles does
not show any definitive trend.

Especially for rectangles A and B, with the most extreme conditions,
we have to be aware of other additional uncertainties, such as $N_{\rm
  {H}}$/\Ejk\ being different than adopted; a decrease in this ratio
(Section~\ref{sec:hydrogen}) would raise both the derived $\sigma_{\rm
  e}(1200)$ and $P$ (deduced from the emission) proportionately, at
constant $T$ (lines in Figure~\ref{fig:usigma}).

Another factor is grain evolution.  
If grains are evolving enough to change the emission opacity
$\sigma_{\rm e}(1200)$ significantly, it is at least plausible that
the capacity to absorb is also changing.  However, there are no
near-infrared to ultraviolet extinction curves from which to quantify
such changes.  The evolutionary effects are important to understand
because the absorption opacity directly affects the power absorbed
from the ISRF and hence $P$ observed in emission.
For the typical sizes of big grains in the diffuse interstellar
medium, the absorption cross section is approaching the geometric
cross section.  With grain growth by accretion and aggregation, this
ratio of cross sections saturates at a value near unity.  Grain growth
also increases the mass faster than the geometric cross section,
driving down the absorption opacity.  Modeling these competing effects
would take a detailed grain model and theory of grain evolution, as
well as an accounting of the spectral shape of the (attenuated) ISRF,
well beyond the scope of this paper.

Another potential factor is the dust-to-gas mass ratio. All of the
fields with data in these figures are at the solar Galactocentric
distance, and so the underlying metallicity is likely the same.
Furthermore, the depletion is already high in the diffuse ISM, leaving
little room for dust mass to increase in more dense regions; however,
ice signatures do appear in molecular clouds.  On the other hand,
there is independent evidence for reduced depletion in some IVCs,
which would lower both $\sigma_{\rm e}(1200)$ and $P$ at constant $T$.
There is a hint of such an effect in Figure~\ref{fig:usigma}.

\subsection{Insight from theoretical modeling of grain evolution}\label{insight}

The emission cross section of dust grains in high latitude diffuse
interstellar clouds is better constrained empirically than in
translucent (A$_{V}$ in the range of 1 to 5 mag) or dense molecular
clouds (Appendix~\ref{literature}).

In their comprehensive review of existing empirical estimates,
\cite{henning1995} comment that ``This large scatter [in submillimeter
  opacities] is very probably not only related to problems with the
observational determination of the opacities but may also reflect real
differences of the dust populations in different environments caused by
evolutionary effects (e.g., coagulation or accretion of mantles).''
This has motivated modeling of the grain evolution and the attendant
changes in opacity.
After reviewing their theoretical calculations relating to the evolution
(ice mantles: \citealp{preibisch1993}; mantles and coagulation:
\citealp{ossenkopf1994}), \cite{henning1995} recommended values for
three notional environments with the opacity increasing in magnitude
from the diffuse interstellar medium (\citealp{draine1984}; like current
estimates) to protostellar cloud envelopes ($n_{\rm H} \sim
10^5$~cm$^{-3}$, \citealp{preibisch1993}; like the values found here for
less dense molecular regions) and even further in dense and cold cores
($n_{\rm H} > 10^7$~cm$^{-3}$, \citealp{ossenkopf1994}; conditions well
beyond what is probed here).
Although it is acknowledged that the models are instructive rather than
definitive, the theoretical estimates have often been adopted for the
analysis of submillimeter data because of the lack of
empirically-calibrated opacities for these environments; the situation
is improving somewhat (Appendix~\ref{literature}) but the most dense and
evolved regions remain challenging.

With deeper targeted surveys of individual clouds using infrared
cameras, some variations of the ratio \Ejh/\Ehk\ have been found,
suggestive of grain evolution.  For example,\footnote{For consistency,
  we have transformed the original photometry to the 2MASS system.}
the ratio found is $1.76 \pm 0.07$ in the $\rho$~Oph cloud
\citep{Kenyon1998}, $1.96 \pm 0.04$ in the Cham~I cloud
\citep{Gomez2001}, and $2.25 \pm 0.07$ in Coalsack Globule~2
\citep{racca2002}.  In these investigations, \Ejk\ ranged up to 5.65,
3.04, and 2.44, respectively, all well beyond the \emph{top} values in
the rectangles considered here.  As emphasized in
Section~\ref{correlation}, \nhoverejk\ has not been calibrated directly
for such high column densities.

The multi-wavelength complexity is highlighted by the recent
calculations by \citet{ormel2011}, motivated by evidence for changes
in the near-infrared extinction for inferred column densities up to
$N_{\rm H} \sim 6 \times10^{22}$~cm$^{-2}$ (notionally \av\ $\sim
32$~mag), again well beyond that probed here.  Following the effects
of ice-mantle formation and (subsequent) grain coagulation, they model
the opacity across the whole spectrum from the ultraviolet to
submillimeter. As with the earlier models, these results show how the
evolution can produce not only increases in submillimeter opacity but
also accompanying dramatic changes in the near-infrared and
ultraviolet opacity.

The combined effects of magnitude and slope changes in the
near-infrared opacity can alter \nhoverejk.  Depending on the model,
significant near-infrared changes might develop even before a
substantial change in submillimeter opacity.  In the Vela region that
we have analyzed above there is no change in the infrared slope, and
assuming no change in the absolute amount of the extinction either
(Section~\ref{sec:hydrogen}) we find that there is an increase in
submillimeter opacity relative to that in the diffuse interstellar
medium. If the magnitude of the infrared opacity, which we cannot
measure directly (but see Figure~\ref{fig:nhvsejk}), has actually
increased, then the derived opacity would be even higher (see
Equation~(\ref{final})).

The \citet{ormel2011} results also show a decrease in
optical-ultraviolet opacity as the grains evolve, a reminder that
$A_V$ might not be a good surrogate of column density in these evolved
regions.  This optical-ultraviolet opacity decrease would decrease the
energy absorbed and thus the observed $P$ in emission.  However,
depending on the details of the evolution, its time development, and
the spectral shape of the radiation field, this decrease might be
compensated by an increase in the near-infrared opacity.  As discussed
above, a mis-calibration of \nhoverejk\ would scale $P$ and
$\sigma_{\rm e}(1200)$ equally, moving the derived quantities along
lines of constant $T$ in Figure~\ref{fig:usigma}.


\section{Conclusion}\label{sec:conclusion}

We have correlated the diffuse interstellar dust emission in the
Galactic plane toward Vela (BLAST images at 250, 350, and
500~\micron\ and the \IRAS\ image at 100~\micron) with a map of
near-infrared color excess made using 2MASS data.  Fourteen regions of
moderate column density were analyzed.  The conversion of color excess
to column density has been examined critically.  From stellar data we
have measured \nh/\Ejk\ to be $11.5 \times 10^{21}$ cm$^2$ mag$^{-1}$
with a considerable dispersion (Figure~\ref{fig:nhvsejk}).  From the
spectral energy distribution of the dust emission, we have quantified
important properties of the big grains, namely the equilibrium
temperature $T$ of the big grains that are in thermal equilibrium with
the interstellar radiation field (ISRF), their submillimeter opacity
$\sigma_{\rm e}(1200)$ (the emission cross section per H nucleon), and
$P$, the total power radiated per H nucleon. We find that:

\begin{enumerate}[1.]

\item The submillimeter opacity is consistently larger than for dust in
  the local high Galactic latitude interstellar medium, by a factor 2 to
  4 relative to the standard $\sigma_{\rm e}(1200) = 1.0 \times
  10^{-25}$~cm$^2$~H$^{-1}$ value ($r\kappa_0 =
  0.043$~cm$^2$~gm$^{-1}$).  This is strong evidence for grain
  evolution.

\item The range of $P$ extends to both higher and lower values compared
  to that found at high latitudes, $3.8 \times 10^{-31}$~W~H$^{-1}$
  (1.2~L$_\odot$/M$_\odot$).  This range in part reflects variations in
  the interstellar radiation field.  It is also influenced by
  evolutionary changes in the dust absorption opacity.  In turn, all of
  the above changes lead to changes in the observable equilibrium $T$.

\item Compared to the local high latitude dust temperature (17.9~K), in
  this direction in the Galactic plane the dust temperatures are
  significantly colder, typically 15~K.  Somewhat lower temperatures
  still are found in more dense higher column density regions where the
  ISRF is more strongly attenuated.

\item Continuing the trend found in high latitude fields, there is an
  inverse correlation of $T$ with $\sigma_{\rm e}(1200)$.  

\end{enumerate}

The recognition that there are changes in the emission opacity raises a
particular point of caution, because the value adopted impacts directly
all column densities deduced from dust emission maps, and the masses of
compact structures (clumps, cores, filaments, ridges) within them.
While values typically being adopted (see Appendix~\ref{literature}) are
within the range that we find, we will need to understand the underlying
causes of the variations already observed in order to assess whether
there are further changes to be expected in the important even denser
environments where the opacity has not been calibrated.



\acknowledgments
  
The BLAST collaboration acknowledges the support of NASA through grant
numbers NAG5-12785, NAG5-13301, and NNGO-6GI11G, the Canadian Space
Agency (CSA), the UK Particle Physics \& Astronomy Research Council
(PPARC), and Canada's Natural Sciences and Engineering Research
Council (NSERC).  We thank the Columbia Scientific Balloon Facility
(CSBF) staff for their outstanding work.  Finally, we appreciate the
careful reading of the manuscript by the referee, B. T. Draine, which
has led to some clarification and elaboration of the analysis.



\appendix

\section{Exploration of the Impact of a $\beta$ -- $T$ Relationship}\label{abeta}

There is a considerable literature on a possible inverse relationship
between $\beta$ and $T$.  However, as is clear from the many examples
in Figure~3 of \citet{paradis2010}, there is no consensus on the
details of this dependence.  In practice, we do not have sufficient
multi-wavelength data to treat $\beta$ as an additional free parameter
in the SED fit; without the fit being well over-constrained, undue
sensitivity could develop to the weighting of the data and issues of
calibration, for example.  Nevertheless, we have gone through the
exercise of treating $\beta$ as a free parameter.  For our rectangles
we found that the values of free-$\beta$ tended to be a bit larger
than 1.8, opposite to the lowering of the apparent $\beta$ expected if
the SED is broadened because of dust of different temperatures
superimposed along the line of sight.

Over all of the regions examined here plus those in
\citet{abergelDd2011} there is a considerable range of $T$ and a
suggestion that $\beta$ is inversely related.  This trend could be
parameterized as
\begin{equation}
\beta/1.8 = (T/17.9\,  {\rm K})^{-2/3}, 
\label{ebeta}
\end{equation}
the power-law index being intermediate among the examples summarized
by \citet{paradis2010}.  

\begin{figure}
\includegraphics[width=\linewidth]{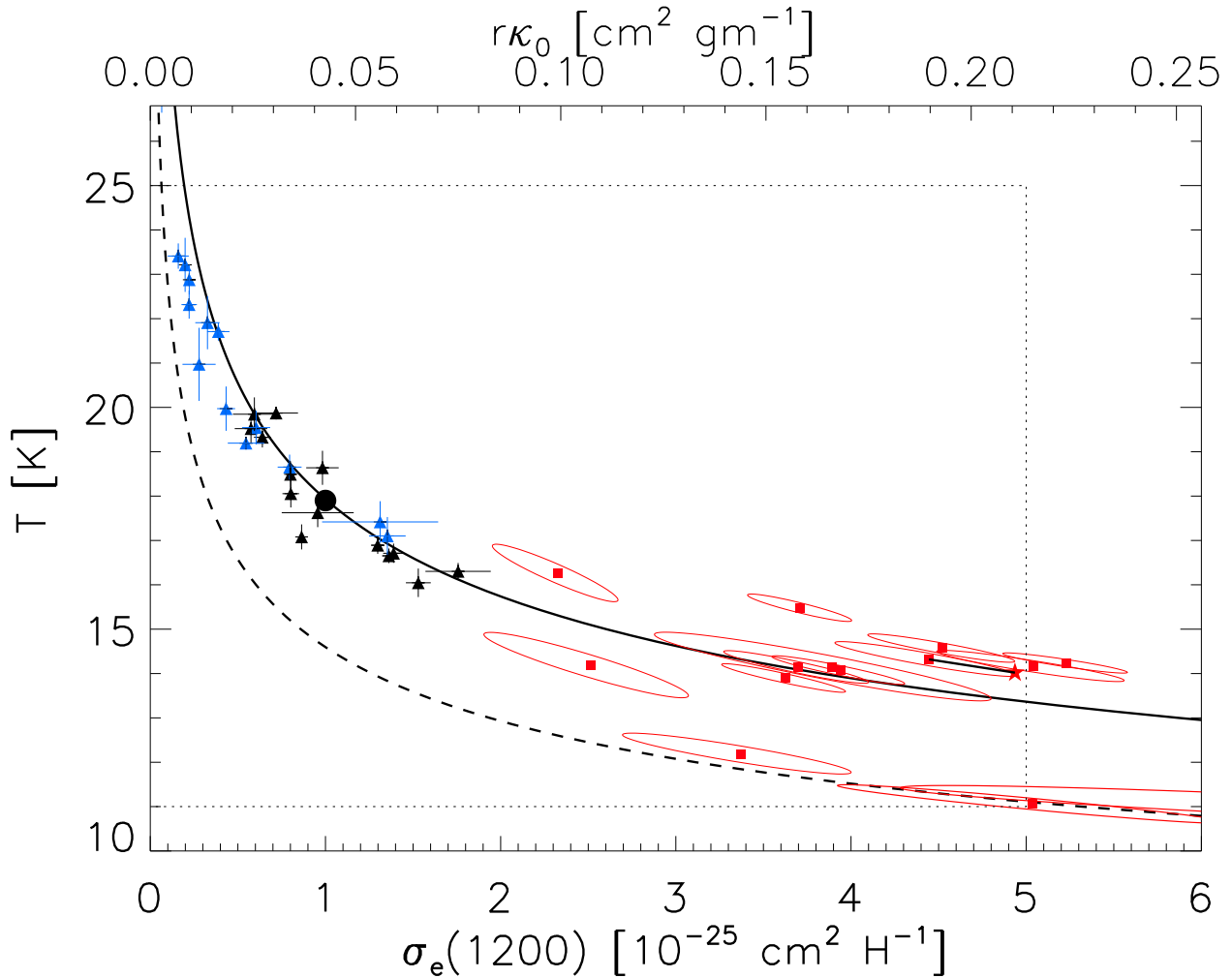} 
\caption{
Dust temperature, $T$, vs.\ emission opacity, $\sigma_{\rm e}(1200)$,
from SED fits under the constraint of the $\beta$ -- $T$ relationship in
Equation~(\ref{ebeta}).
Loci are for constant $P$ under the same constraint, for the values
used in Figure~\ref{fig:tsigma}.
Compared to Figure~\ref{fig:tsigma}, the solutions for the parameters
are spread out over a larger range about the fiducial values
corresponding to $\beta = 1.8$ and $T_0=17.9$~K.  However, the
increased spread is systematic, each point remaining near the same
value of $P$ found for fixed $\beta$.
Because of this spreading, the coverage of this figure has been
expanded relative to that of Figure~\ref{fig:tsigma} (dotted
outline) and even so the most extreme points, an IVC and rectangle B,
lie off the plot to the upper left and lower right, respectively.
For rectangle~\rectno, the star represents the solution obtained when
$\beta$ is treated as a free parameter in the SED fit; $\beta$ is
found to be 2.2, slightly above the trend described by
Equation~(\ref{ebeta}), and so that solution is similar to the
constrained result here.}
\label{fbeta}
\end{figure}

For the purposes of illustration, following this possible trend
through to its potential consequences, we have performed SED fits
subject to the added constraint of Equation~(\ref{ebeta}).  The
derived parameters are displayed in Figure~\ref{fbeta}.  Loci for
constant $P$ under the same constraint are plotted for reference.
The effect of constraining the SED fit with this $\beta$ -- $T$
relationship is systematic.  As expected, each solution remains near
the same value of $P$ found for fixed $\beta$.  However, compared to
Figure~\ref{fig:tsigma}, the values of the other two parameters are
spread out over a larger range on either side of the fiducial values
corresponding to $\beta = 1.8$ and $T_0=17.9$~K.

This spreading exaggerates the differences in $\sigma_{\rm e}(1200)$
found from region to region using fixed $\beta$.
We are not actually persuaded that there is a $\beta$ -- $T$
relationship, but in any case we conclude that the estimated magnitude
of the environmental variations as found for fixed $\beta = 1.8$ and
discussed in the paper is conservative.

\section{Values of Opacity in the Literature}\label{literature}

As in Section~\ref{dustemission}, we standardize on 1200~GHz (250~\micron)
as the fiducial reference frequency $\nu_0$, a value quite appropriate
to on-going analyses of \emph{Herschel} data, for example.  The opacity
$\sigma_{\rm e} (1200)$ and the product $r \kappa_0$ can be used
interchangeably through Equation~(\ref{cross}), as in the lower and upper
axes in Figures~\ref{fig:tsigma} and \ref{fig:usigma}; the scale factor
between the respective numbers on the axes is 0.043.
Scaling to and from the value of opacity at another frequency depends on
$\beta$ in which there is some uncertainty; we use $\beta = 1.8$.  For
example, scaling to 1000~GHz lowers the opacity by a factor 1.4.

\subsection{Adopted}\label{adopted}

A value representative of the \citet{preibisch1993} theoretical
core-mantle grain evolution results for conditions in the cloud
envelopes of prestellar cores, corresponding to 0.1~cm$^2$~gm$^{-1}$ at
1000~GHz or $r \kappa_0 = 0.14$~cm$^2$~gm$^{-1}$, has been adopted by
the \emph{Herschel} Guaranteed Time Key Programs on star formation
(Gould Belt Survey, \citealp{andre2010}; HOBYS, \citealp{motte2010}),
allowing consistent comparisons between different regions analyzed.
Whether this is valid for the dense cores extracted is uncertain.

Analyses based on data (only) at lower frequencies, for example by
\citet{kerton2001} for data from SCUBA at JCMT and by \cite{motte2007}
for data from MAMBO at IRAM, often adopt an opacity at a lower fiducial
frequency.  These are based on theoretical estimates for $n_{\rm H} \sim
10^5$~cm$^{-3}$ \citep{ossenkopf1994}, and would scale compatibly using
to $\beta = 2$ to $r \kappa_0 = 0.23$~cm$^2$~gm$^{-1}$ (or
0.18~cm$^2$~gm$^{-1}$ for $\beta = 1.8$).  This is very similar to the
value adopted for \emph{Herschel} because of the closely related
theoretical basis.
%

%
Analyzing the compact sources in the Science Demonstration Phase (SDP)
fields of the Open Time Key Program Hi-GAL \citep{molinari2010},
\citet{elia2010} adopted not only a low fiducial frequency (230~GHz) but
also a variable $\beta$ in fitting the SEDs; thus comparisons of derived
masses are not so straightforward, although effects of $\beta$ on $T$
and the implied $r \kappa_0$ somewhat cancel in the mass estimates.

In most BLAST papers on Galactic star forming regions \citep{chapin2008,
  truch2008, rivera2010, roy20115146, roy2011cygx} we adopted $r
\kappa_0 = 0.10$~cm$^2$~gm$^{-1}$ (\citealp{hildebrand1983}; derived
empirically from limited observations of a single molecular cloud, for
which the value is ``probably good within a factor three or four,'' this
has nevertheless often been adopted as a ``canonical'' value). 
In the BLAST analysis of the Vela molecular ridge region
\citep{netterfield2009, luca2009} we estimated 0.16~cm$^2$~gm$^{-1}$
using a CO calibration (Section~\ref{opacity}).
For the diffuse interstellar medium toward the CasA supernova remnant
\citep{sibthorpe2010} we used 0.05~cm$^2$~gm$^{-1}$.  A low value like
this is also implicit in any analysis with the standard grains in DustEM
\citep{compiegne2011}.

\subsection{Empirical}\label{empirical}

The value of the opacity for the diffuse high latitude interstellar
medium is the best determined, through correlations of the dust emission
with \nh\ from observations of the 21~cm emission line.
Using observations from DIRBE and FIRAS, \cite{boulanger1996} obtained
an emission cross-section $\sigma_{\rm e} (1200) = 1\, \times
10^{-25}$~cm$^2$~H$^{-1}$, and an equilibrium temperature of 17.5~K, in
agreement with the value obtained by \cite{draine1984}.  Also using
\emph{COBE} data, \cite{lagache1999} found $0.87\pm 0.09 \times
10^{-25}$~cm$^2$~H$^{-1}$, compatible also with the result obtained by
\cite{weingartner2001}.  In the text here we cited and plotted $r
\kappa_0 = 0.043$~cm$^2$~gm$^{-1}$ for $T = 17.9$~K
\citep{abergelDd2011}.
Recently, multi-wavelength analysis of correlations of \emph{Planck} and
\emph{IRAS} dust maps covering submillimeter to far-infrared wavelengths
\citep{abergelDd2011} with new higher-resolution observations of
\nh\ from the GBT (\citealp{boothroyd2011}, Martin et al.\ in
preparation) showed that there were regional variations about this value
(see Figure~\ref{fig:tsigma}).
The atomic hydrogen column density associated with these high latitude
clouds ranges over $N_{\rm H} \sim 0.2$ to $6
\times10^{20}$~cm$^{-2}$ which is equivalent to \av\ $\sim 0.01$ to
0.3~mag.
For the atomic phase in the Taurus field, \citet{abergelTd2011} obtained
a similar opacity $1.14 \pm 0.2 \times 10^{-25}$~cm$^2$~H$^{-1}$ using
\emph{Planck} data, for column densities up to $30
\times10^{20}$~cm$^{-2}$ (\av\ about 1.6~mag).

In their Section~5.4.1, \citet{abergelTd2011} review earlier estimates
of opacity in more molecular regions (see also Section~5.2 in
\citealp{ juvela2011} and Figure~3b in \citealp{Kramer2003}).  For the
molecular phase in the same Taurus field they obtained a higher
opacity, $2.3 \pm 0.3 \times 10^{-25}$~cm$^2$~H$^{-1},$ using
\emph{Planck} data and gauging the column density using 21~cm and CO
emission-line observations ($N_{\rm H} \sim 1\,
\times10^{22}$~cm$^{-2}$).  Values for both the atomic phase and
molecular phase are plotted in Figure~\ref{fig:tsigma}; they follow
basically the same trend established by the other data, on a locus of
slightly lower power (see also Figure~\ref{fig:usigma}).

Also in the Taurus region, \cite{terebey2009} find a similarly higher
opacity, by correlating dust optical depth ($T = 14.2$~K) from MIPS
imaging from \emph{Spitzer}, extending to 160~\micron, with the column
density.  To estimate the column density they used near-infrared
color-excess (which has the same uncertainties as elaborated here in
Section~\ref{sec:hydrogen}), making a careful analysis of \av.
Transforming their 160-\micron\ opacity to our fiducial frequency, we
find $\sigma_{\rm e} (1200) = 2.1 \times 10^{-25}$~cm$^2$~H$^{-1}$ for
$\beta = 2$ (2.3 for $\beta = 1.8$); see Figures~\ref{fig:tsigma} and
\ref{fig:usigma}.  \citet{flagey2009} also analyzed similar data over a
slightly bigger map in Taurus, finding a very similar temperature and
opacity.

With the advent of submillimeter mapping by \emph{Planck} and
\emph{Herschel} (with zero-point offsets from \emph{Planck}), SEDs
have been fit pixel by pixel, resulting in maps of dust optical depth
$\tau$ and $T$.  From the slope of $\tau$ versus \nh\ estimated from
observations of \ion{H}{1}\ and CO, \citet{bernard2010} find values $r
\kappa_0 = 0.094$ and 0.14~cm$^2$~gm$^{-1}$ for the Hi-GAL SDP fields
at $\ell = 59$\degree\ and $\ell = 30$\degree, respectively.

Dividing the $\tau$ map directly by a map of column density produces a
map of opacity (Equation~(\ref{cross})).  Note that for this
application column density is often obtained by converting a
near-infrared color excess (unnecessarily expressed as \av) into \nh\
and so the very same caution (Sections~\ref{sec:hydrogen} and
\ref{insight}) as to the lack of direct calibration at high column
density applies, even more so.  This pixel by pixel approach has the
advantage of tracking opacity changes at higher spatial resolution
compared to the correlation analyses used here.  These spatial changes
can be related to changes in $T$ as well, e.g., producing a
scatter-plot version of Figure~\ref{fig:tsigma}.  However, it is worth
recalling that this method essentially assumes that the properties of
the dust ($T$, opacity) are uniform along the line of sight, which
might not be the case when the properties are apparently changing
significantly in the transverse direction.
Mapping the environs of \emph{Planck} cold clumps with \emph{Herschel},
\citet{juvela2011} find opacities typically 0.1~cm$^2$~gm$^{-1}$ where
$T \sim 17$~K ($\beta = 2$), but that on the high column density lines
of sight $T$ drops to 14.5~K and the opacity rises to 0.2 to
0.3~cm$^2$~gm$^{-1}$.  This again supports the trend found in
Figure~\ref{fig:tsigma}, now on a locus of slightly larger power.






\end{document}